\documentclass[5p,10pt]{elsarticle}       

\usepackage[utf8]{inputenc}
\usepackage{amsfonts}
\usepackage{amsmath}
\usepackage{multicol}
\usepackage{amssymb}
\usepackage{graphicx}
\usepackage{mathtools}
\usepackage{xcolor}
\usepackage{blindtext}

\usepackage{colortbl}
\usepackage{arydshln}
\usepackage{booktabs}
\usepackage{hhline}

\usepackage{breqn}
\usepackage{enumitem}
\newcommand{\subscript}[2]{$#1 _ #2$}

 \usepackage{breqn}

\usepackage{units}

\newcounter{theExample}
\newenvironment{Example}{
        \par\smallskip\small\refstepcounter{theExample}%
        \noindent\hspace*{-0em}\textbf{Example~\arabic{theExample}}:~%
        \leftskip0em\ignorespaces%
}{
        \par\smallskip
}
\newcounter{theRemark}
\newenvironment{Remark}{
        \par\smallskip\refstepcounter{theRemark}%
        \noindent%
        \textbf{Remark~\arabic{theRemark}}:~%
        \ignorespaces%
}{
        \par\smallskip
}

\begin{document}
        
\title{Vertex-Frequency Graph Signal Processing: A Review}

\author[etf]{Ljubi\v{s}a Stankovi\'{c}}
\ead{ljubisa@ucg.ac.me}
\author[ic]{Danilo Mandic}
\ead{d.mandic@imperial.ac.uk}
\author[etf]{Milo\v{s} Dakovi\'{c}} 
\ead{milos@ucg.ac.me} 
\author[ic]{\\Bruno Scalzo}
\ead{bruno.scalzo-dees12@imperial.ac.uk}
\author[etf]{Milo\v{s} Brajovi\'{c}}
\ead{milosb@ucg.ac.me}
\author[pit]{\!Ervin \!Sejdi\'{c}}
\ead{esejdic@pitt.edu}
\author[ic]{Anthony G. Constantinides}
\ead{a.constantinides@imperial.ac.uk}

\address[etf]{University of Montenegro, Podgorica, Montenegro }
\address[ic]{Imperial College London, London, United Kingdom }

\begin{abstract}
Graph signal processing deals with signals which are observed on an irregular graph domain. While many approaches have been developed in classical
graph theory to cluster vertices and segment large graphs in a signal independent way,  signal localization based approaches to the analysis of data on graph represent a new research direction which is also a key to big data analytics on graphs. To this end, after an overview of the basic definitions in graphs and graph signals, we present and discuss a localized form of the graph Fourier transform. To establish analogy with classical signal processing, spectral- and vertex-domain definitions of the localization window are given next. The spectral and vertex
localization kernels are then related to the wavelet transform, followed by their polynomial approximations and a study of filtering and inversion. For rigor, the analysis of energy representation and frames in the localized graph Fourier transform is extended to the energy forms of vertex-frequency distributions, which operate even without the need to apply localization windows.
Another link with classical signal processing is established through the concept of local smoothness, which is subsequently related to the particular paradigm of signal smoothness on graphs. This all represents a comprehensive account of the relation of general vertex-frequency analysis with classical time-frequency analysis, an important but missing link for more advanced applications of graphs signal processing. The
theory is supported by illustrative and practically relevant examples.    
\end{abstract}
\maketitle
%\tableofcontents

\setcounter{tocdepth}{3}

\section{Introduction}

Processing of data whose domain is a graph attracted significant research interest recently \cite{sandryhaila2013discrete,chen2015discrete,sandryhaila2014discrete,ortega2018graph,hamon2016extraction,Tutorialicassp2017,quinn2015directed,raginsky2011performance,hamon2019transformation}. The Big Data paradigm has revealed the possibility of using smaller and localized subsets of the available information to enable a reliable mathematical analysis and local characterization of subsets of data of interest \cite{sandryhaila2014big}. Oftentimes in practical applications concerned with large graphs, we may not be interested in the analysis 
of the entire graph signal, but rather in its local behavior. Aiming to characterize the localized signal behavior in the joint vertex-frequency domain, a natural analogy with the classical time-frequency analysis is established \cite{stankovic2014time,cohen1995time,boashash2015time}. 

Indeed, the concept of signal localization by using window functions has been extended 
to signals defined on graphs \cite{shuman2016vertex, shuman2012windowed,zheng2016multi,tepper2016short,stankovic2017vertex,cioacua2019graph,hammond2019spectral,behjat2019spectral}. This 
extension is not straightforward, since, owing to inherent properties of graphs which are irregular but interconnected domains, even an operation which is very simple in classical time-domain analysis, like the time shift, 
cannot be straightforwardly generalized to graph signal domain. This has resulted in several approaches 
to the definition of the graph shift operator, and much ongoing research in this domain.

A common approach to windowing in the graph domain is to utilize the signal spectrum to obtain  window functions 
for each graph vertex \cite{shuman2013emerging}. Another possibility is to define a 
window support as a local neighborhood for each vertex \cite{stankovic2017vertex}.
In either case, the localization window is defined by a set of vertices that contain the 
current vertex, $n$, and all vertices that are close to the vertex $n$, that is, a neighborhood of vertex $n$. Special attention is paid to the local graph Fourier transform approaches that can be implemented in the vertex domain, since this domain can be a basis for numerically efficient analysis in the case of very large graphs. 

As in the 
classical signal analysis, a localization window should be narrow enough in order to provide 
good localization of the signal properties but wide enough to produce high 
resolution in the spectral domain. To automatize the process of making this compromise, optimization approaches shall be involved, some of which are related to the uncertainty principle.  

Vertex-frequency analysis can be used for graph signal estimation, filtering, and efficient representation. In many of these applications the graph signal should be reconstructed after an appropriate processing in the vertex-frequency domain. Two forms of the local graph Fourier transform inversion are considered. The inversion condition is defined within the frames framework as well. The local graph Fourier transform implementation and its inversion are related to the graph wavelet transform. 

Finally, the energy versions of the vertex-frequency representations are considered. These representations can be implemented without a localization window. They are studied as the local smoothness index estimators. The reduced interference vertex-frequency distributions, which satisfy the marginal property and localize graph signal energy in the vertex-frequency domain are defined and related to the classical time-frequency analysis, as a special case. All concepts are illustrated through examples. 

The paper is organized as follows. The basic definitions of graph signals and spectral graph domain are given in Section II. The localized graph Fourier transform is presented in Section III, where various approaches to define this transform are considered. The local graph Fourier transform is also related to the wavelet transform. The topic of Section IV is the optimization of the graph signal localization window, while in Section V, the inversion relations and conditions for the considered graph transforms are given. The uncertainty principle in graph signals is reviewed in Section VI. The graph spectrogram is related to the frames in Section VII. The energy vertex-frequency representations are defined and analyzed in Section VIII. The paper ends with concluding remarks and a reference list.

\section{Review of Basic Definitions}

Consider a graph with $N$ vertices, denoted as $n \in\mathcal{V}=\{1,2,\dots, N\}$, connected with edges whose weights are $W_{mn}$.  If the vertices $m$ and $n$ are not connected then $W_{mn}=0$. The edge weights $W_{mn}$ are commonly written in an $N\times N$ matrix form, $\mathbf{W}$.  For undirected graphs the weight matrix $\mathbf{W}$ is symmetric, $\mathbf{W}=\mathbf{W}^T$. In the case of unweighted, graphs all nonzero elements in $\mathbf{W}$ are equal to unity, and this specific form is  called the connectivity or adjacency matrix, denoted by $\mathbf{A}$.

For an enhanced graph description, in addition to matrices $\mathbf{A}$ or $\mathbf{W}$, it is common to use a diagonal degree matrix $\mathbf{D}$, whose elements $D_{nn}$ are equal to the sum of all edge weights connected to the considered vertex, $n$, that is, $D_{nn}=\sum_m W_{mn}$, and indicate the vertex importance.  The weight matrix and the degree matrix can be combined to produce the graph Laplacian, given by $\mathbf{L}=\mathbf{D}-\mathbf{W}$. The Laplacian of an undirected graph is symmetric, $\mathbf{L}=\mathbf{L}^T$. 

Spectral analysis of graphs is most commonly based on the eigendecomposition of the graph Laplacian $\mathbf{L}$ or the adjacency matrix, $\mathbf{A}$. Some of the most common eigendecompositions in graph signal analysis are given in Table \ref{tab:1a}. By default, we will assume the decomposition of the graph Laplacian $\mathbf{L}$, if not stated otherwise in the paper. The eigenvectors, $\mathbf{u}_k$, and the eigenvalues, $\lambda_k$, of the graph Laplacian are calculated based on the usual definition $\mathbf{L}\mathbf{u}_k=\lambda_k \mathbf{u}_k$, $k=1,2,\dots,N$.

 \begin{table}[tb]
	
	\centering
		\caption{Summary of graph spectral basis vectors.  }
	
	\renewcommand{\arraystretch}{1.25}
	\begin{tabular}{|l|l|}
		\hline\hline
		\textbf{Operator} & \textbf{Eigenanalysis relation}  \\[5pt]
		\hline\hline 
		Graph Laplacian & $\mathbf{L}\mathbf{u}_k=\lambda_k\mathbf{u}_k$  \\[5pt]
		\hline\hline
		Generalized eigenvectors    &   \\[0pt] 
		of graph Laplacian  &  $\mathbf{L}\mathbf{u}_k=\lambda_k\mathbf{D}\mathbf{u}_k$ \\[5pt]
		\hline\hline
		Normalized graph  Laplacian & $\big(\mathbf{D}^{-1/2}\mathbf{L}\mathbf{D}^{-1/2}\big)\mathbf{u}_k=\lambda_k\mathbf{u}_k$ \\[5pt]
		\hline\hline 
		Adjacency matrix & $\mathbf{A}\mathbf{u}_k=\lambda_k\mathbf{u}_k$ \\[5pt]
		\hline\hline 
		Normalized adjacency matrix & $\Big(\frac{1}{\lambda_{\max}}\mathbf{A}\Big)\mathbf{u}_k=\lambda_k\mathbf{u}_k$  \\[5pt]
		\hline\hline
	\end{tabular}
	\renewcommand{\arraystretch}{1}
	\label{tab:1a}	
\end{table}

The graph Fourier transform (GFT), $$\mathbf{X}=[ X(1),X(2),\allowbreak \dots , X(N) ] ^T,$$ of a graph signal, $\mathbf{x}=[ x(1),x(2),\allowbreak \dots , x(N) ] ^T$,  is defined an expansion onto a set of orthonormal basis functions, $\mathbf{u}_k$, the elements of which are $u_k(n)$, $n=1,2,\dots,N$, that is 
 \begin{equation}
 X(k)=\mathrm{GFT}\{x(n)\}=\sum_{n=1}^N x(n)\; u_{k}(n). \label{GFT}
 \end{equation}
 The inverse graph Fourier transform (IGFT) is then defined as 
 \begin{equation}
x(n)=\mathrm{IGFT}\{X(k)\}=\sum_{k=1}^N X(k)\; u_{k}(n). \label{IGFT}
\end{equation}
This set of transforms reduces to the classical pair of the discrete Fourier transform (DFT) and the inverse DFT, if the graph is circular and directed. The eigenvectors of a directed graph are complex-valued and for this case we should use complex-conjugate basis functions, $u^*_{k}(n)$, in (\ref{GFT}).

\section{Localized Graph Fourier Transform (LGFT)}\label{s:VLS}

The localized graph Fourier transform (LGFT) can be considered an extension of the standard localized time (short time) Fourier transform (STFT), and can be  calculated as the GFT
of a signal, $x(n)$, multiplied by an appropriate vertex localization window function, $h_m(n)$, to yield
\begin{equation}
S(m,k)=\sum_{n=1}^N x(n)h_m(n)\; u_{k}(n). \label{VFSPEC}
\end{equation}
In general, is assumed that the graph window function, denoted by $h_m(n)$, should be such that it 
localizes the signal content around the vertex $m$. Its values should be close to $1$ at vertex $m$ and vertices in its close neighborhood, and should tend to $0$ for vertices far from  vertex $m$.  %For an illustration of the localization window on a graph see Fig. \ref{VF_windows_LVS}, panels (a) and (c).

The local GFT in a matrix notation, $\mathbf{S}$, contains all its elements, $S(m,k)$, $m=1,2,\dots,N$, $k=1,2,\dots,N$. The columns of $\mathbf{S}$ which correspond to a vertex $m$ are given by 
$$\mathbf{s}_m=\mathrm{GFT}\{x(n)h_m(n)\}= \mathbf{U}^T\mathbf{x}_{m},$$ where  $\mathbf{x}_{m}$ is the vector whose elements $x(n)h_m(n)$ are equal to the graph signal samples, $x(n)$, multiplied by the window function, $h_m(n)$, centered at the vertex $m$, and matrix $\mathbf{U}$ is composed of eigenvectors $\mathbf{u}_k,~k=1,2,\dots,N$ as its columns. 

In general the set of vertices, $m$, and spectral indices, $k$, where the LGFT is calculated may be reduced from the original number of $N^2$ vertex-frequency points, $(m,k)$. For a nonredundant realization, $N$ points and corresponding values, $S(m.k)$, would be sufficient \cite{stankovic2015stft}.

\smallskip
\noindent\textbf{Special cases:}
\begin{itemize}
\item 
For $h_m(n)=1$, the localized vertex spectrum is  equal to the standard 
spectrum 
$S(m,k)=X(k)$ in (\ref{GFT}) for each $m$, meaning that no vertex localization is 
performed. 
\item 
If $h_m(m)=1$ and $h_m(n)=0$ for $n\ne m$, the localized vertex 
spectrum is equal to the signal $S(m,0)=x(m)/\sqrt{N}$ for $k=0$. 
\end{itemize}

The subsequent subsection outlines ways to create windows in the vertex domain.  We shall address two methods for defining graph localization window functions $h_m(n)$. The analysis will first  focus on the windows, $h_m(n)$, defined using their basic function in the spectral domain. Next, the spectral domain definition will be related  to the wavelet transform. Subsequently, a  purely vertex domain window formulation will be presented.

\subsection{Windows Defined in the GFT Domain}

\noindent\textbf{Generalized convolution of graph signals.} Consider two signals, $x(n)$ and 
$y(n)$,  defined on a graph. The corresponding  GFTs  are denoted by $X(k)$ and $Y(k)$. \textit{A
generalized convolution}, $z(n)$, of signals $x(n)$ and $y(n)$ can then be defined in the GFT domain, in analogy with the classical definition of convolution, as
\begin{gather}
z(n)  =x(n) * y(n) = \mathrm{IGFT}\{Z(k)\}, \text{ where } \nonumber \\
Z(k) =\mathrm{GFT}\{x(n) * y(n)\} = X(k)Y(k). \label{GdnClnvG}
\end{gather}

\noindent\textbf{Shift operator for graph signals.} A ``shift'' on a graph cannot be extended  to graph signals in a direct analogy to the 
classical signal shift. Among several forms of signal shift on a graph which have been proposed in graph theory, the most popular graph shift operator is based on the multiplication of the signal with the adjacency matrix; here we will use the definition of shift operator based on a generalized convolution \cite{shuman2016vertex}.
 Consider the graph signal,
$h(n)$, and the delta function located at a vertex $m$, given by $\delta_m(n)= \delta(n-m)$. The  GFT of delta function, $\delta_m(n)$, is then given by
$$\mathrm{\Delta}_m(k)=\mathrm{GFT}\{\delta_m(n)\}=\sum_{n=1}^N \delta_m(n) u_k(n) =u_k(m).$$

We will use the symbol $h_m(n)$
to denote a shifted version of the graph signal, $h(n)$, \textquotedblleft toward\textquotedblright \, a vertex $m$. Based on (\ref{GdnClnvG}), this kind of graph signal shift will be defined following the same reasoning as in classical signal processing, where the shifted signal is obtained as a convolution of the original signal and an appropriately shifted delta function. Therefore, a graph shifted signal can be defined as a generalized graph convolution, $h(n)*\delta_m(n)$,  the GFT of which is equal to 
\begin{equation}
\mathrm{GFT}\{h(n)*\delta_m(n)\} = H(k)u_k(m). \label{GSHVER}
\end{equation}
The graph-shifted signal then represents the IGFT of $H(k)u_k(m)$, so that from (\ref{GdnClnvG}) the window localized at the vertex $m$, denoted by $h_m(n)$, is given by \cite{shuman2012windowed}
\begin{equation}
h_m(n)=h(n)*\delta_m(n)=\sum_{k=1}^NH(k)u_k(m)u_k(n).
\label{win-spect1}
\end{equation}
The basic function of this window, $h(n)$, can be conveniently defined in the spectral domain, for example, in the form
\begin{equation}
H(k)=C
\exp (-\lambda_k \tau),
\label{win-spect}
\end{equation}
where $C$ denotes the window amplitude and $\tau>0$ is a 
constant which determines the window width. An example of two windows obtained in this way is illustrated in Fig. \ref{VF_windows_LVS}(a), (b). Observe that the exponential function in (\ref{win-spect}) corresponds to a Gaussian window in  classical analysis (thus offering the best time-frequency concentration \cite{stankovic2014time,cohen1995time,boashash2015time}),  since  graph signal processing on a path graph reduces to classical signal analysis. In that case, the eigenvalues of the graph Laplacian, $\lambda$, may be related to the frequency, $\omega$, in the classical signal analysis as $\lambda \sim \omega^2$. 

\noindent\textbf{ Properties of graph window functions.} The  window localized at the vertex $m$, and defined by (\ref{win-spect1}), satisfies the following properties:
\begin{enumerate}[label=\subscript{W}{{\arabic*}}:] 
        \item Symmetry, $h_m(n)=h_n(m)$. This property follows from the definition in (\ref{win-spect1}), for real-valued basis functions.
        \item A sum of all localized window coefficients, $h_m(n)$, is equal to $H(1)$, since
        \begin{gather*}
        \sum_{n=1}^Nh_m(n)=\sum_{k=1}^NH(k)u_k(m)\sum_{n=1}^Nu_k(n)\\
        =
        \sum_{k=1}^NH(k)u_k(m)\delta(k-1)\sqrt{N}=H(1),
        \end{gather*}
with $\sum_{n=1}^Nu_k(n)=\delta(k-1)\sqrt{N}$, by definition of the eigenvectors, $u_k(n)$.
        \item
        The Parseval theorem for  $h_m(n)$ has the form
        \begin{equation}
        \sum_{n=1}^N|h_m(n)|^2=\sum_{k=1}^N|H(k)u_k(m)|^2.
        \end{equation}
\end{enumerate}
These properties  will be used in the sequel in the inversion analysis of the LGFT.

For the window form in (\ref{win-spect1}), the LGFT can be written as 
\begin{gather}
S(m,k)=\sum_{n=1}^N x(n)h_m(n)\; u_{k}(n) \label{LGFTDEF}\\
=\sum_{n=1}^N \sum_{p=1}^Nx(n)H(p)u_p(m)u_p(n) \; u_{k}(n). \label{LGFTDef1}
\end{gather}
The modulated (spectral shifted) version of the window centered at vertex $m$ and for spectral index $k$ will be referred to as the \textit{vertex-frequency kernel}, $\mathcal{H}_{m,k}(n)$, defined as
\begin{equation}
\!\mathcal{H}_{m,k}(n)\!\!=\!\!h_m(n)u_k(n)\!\!=\!\!\Big(\! \sum_{p=1}^N\!\!H(p)u_p(m)u_p(n)  \Big) u_k(n). \label{KernelGFDFTmk}
\end{equation}
Using the kernel notation, it becomes obvious that the LGFT, for a given vertex $m$ and spectral index $k$, physically represents  \textit{a projection of a graph signal $x(n)$ onto the graph kernel} $\mathcal{H}_{m,k}(n)$, that is,
\begin{equation}
S(m,k)=\langle\mathcal{H}_{m,k}(n),x(n)\rangle=\sum_{n=1}^N\mathcal{H}_{m,k}(n)x(n).\label{KernelGFDFTDEF}
\end{equation}

\begin{Remark}
	The classical STFT, a basic tool in time-frequency analysis, can be obtained as a special case of the GFT when the graph is directed and circular. For this type of graph, the adjacency matrix decomposition produces complex-valued eigenvectors of the form $u_k(n)\sqrt{N} =\exp(j 2 \pi (n-1)(k-1)/N)$, $n=1,2,\dots,N$,   $k=1,2,\dots,N$. Then, having in mind the complex nature of eigenvectors, the value of $S(m,k)$ in (\ref{LGFTDEF})  becomes the standard STFT, that is
	\begin{align}
S(m,k)&=\frac{1}{N^{3/2}}\sum_{n=1}^{N} \sum_{p=1}^{N}x(n)H(p)e^{-j\frac{2 \pi}{N} (m-1)(p-1)}\notag\\
& ~~~~~~~~~~~~~~~~ \times e^{j  \frac{2 \pi}{N} (n-1)(p-1)}e^{-j  \frac{2 \pi}{N} (n-1)(k-1)} \nonumber \\
&=\frac{1}{N}\sum_{n=1}^{N} x(n)h(n-m)e^{-j 2 \pi (n-1)(k-1)/N},
\end{align}
	where $h(n)$ is the inverse DFT of $H(k)$.
\end{Remark}

\begin{figure*}
	\centering
	
	\includegraphics[scale=0.9]{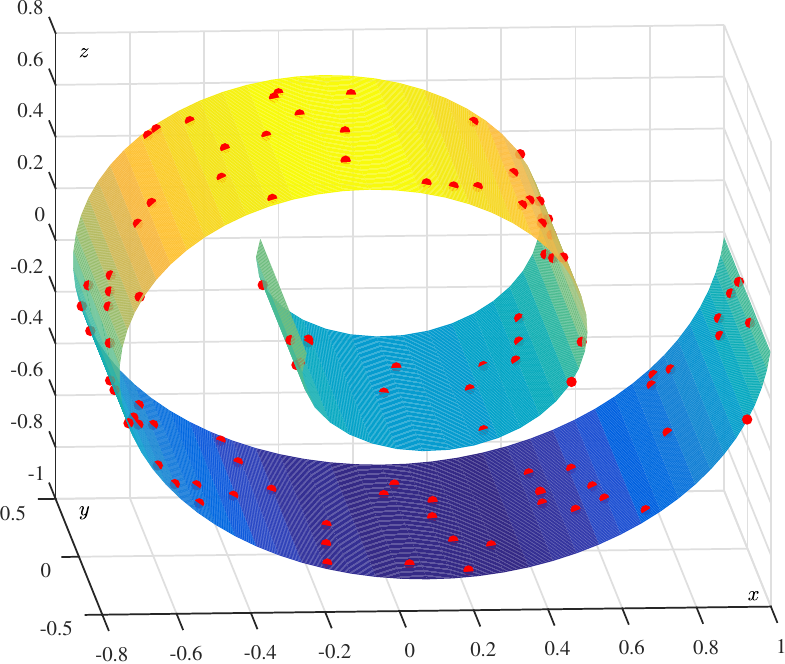}\hspace{2mm}(a)\hspace{5mm}
	\includegraphics[scale=0.9]{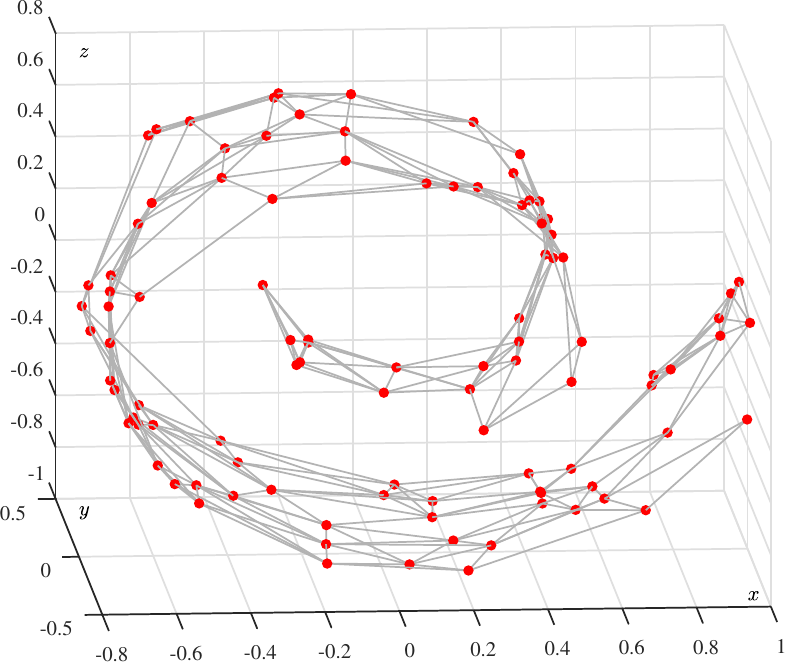}\hspace{2mm}(b)
	
	\vspace{10mm}
	
	\includegraphics[scale=0.9]{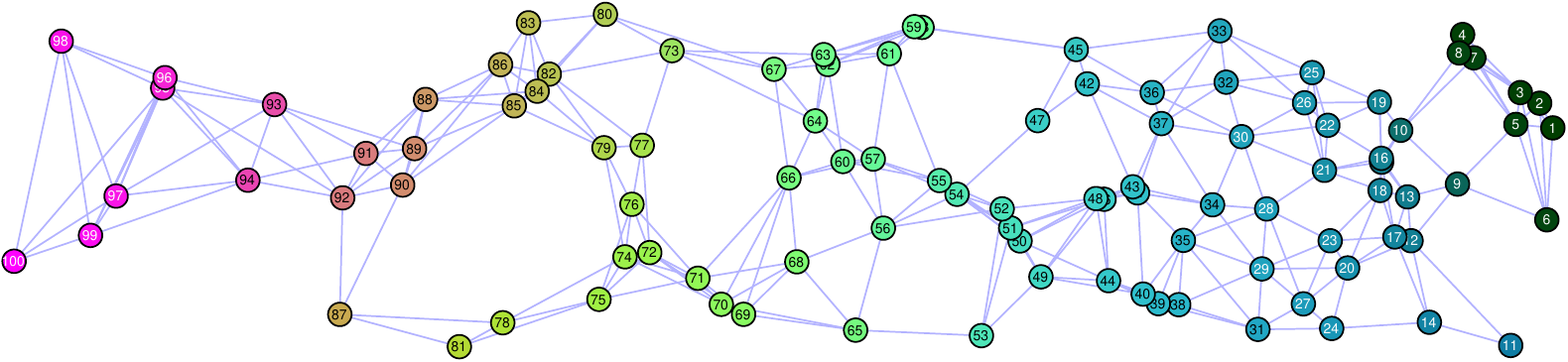}\hspace{7mm}(c)
	
	\vspace{5mm}
	
	\includegraphics[scale=0.9]{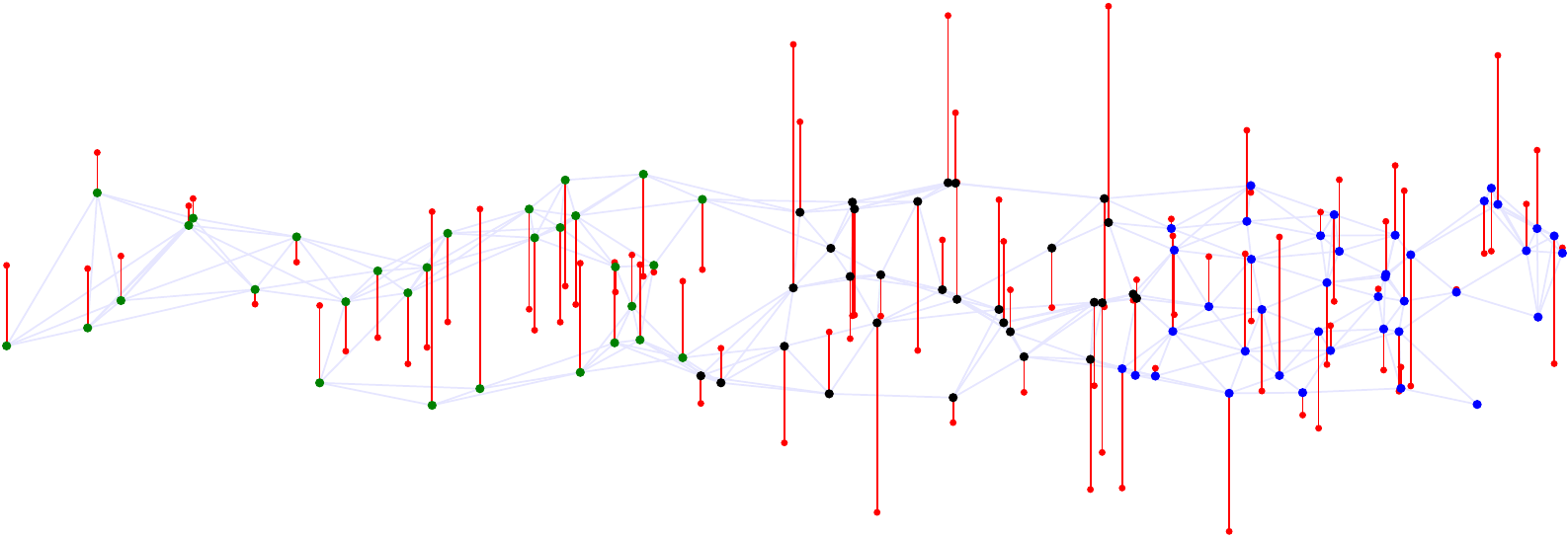}\hspace{7mm}(d)
	
	\caption{Concept of a graph and a signal on a graph. (a) Vertices on a three-dimensional  Swiss roll surface. (b) A graph representation on the Swiss roll manifold. (c) Two-dimensional presentation of the three-dimensional graph from (b), with vertex colors defined by the graph Laplacian eigenvectors $u_1(n)$, $u_2(n)$, and $u_3(n)$. (d) A signal on the graph in (c), which is composed of three eigenvectors (signal components). Supports of these three components  are designated by different vertex colors. The vertex-frequency representations are then assessed based on their ability to clearly resolve and localize these three graph signal components. }
	\label{VF_ex1ab}
\end{figure*}

        \begin{Example} To illustrate the principle of  local vertex-frequency representation consider the graph and the graph signal from Fig. \ref{VF_ex1ab}.
        	A graph with $N=100$ vertices, randomly placed on the so called Swiss roll surface, is shown in Fig. \ref{VF_ex1ab}(a). The vertices are connected with edges whose weights are defined as $W_{mn}=\exp(-r_{mn}^2/\alpha)$, where $r_{mn}$ is the Euclidean  distance between vertices $m$ and $n$, measured on the Swiss roll manifold, and $\alpha$ is a constant. The small weight values are hard-thresholded to zero, to reduce the number of edges associated with each vertex to only a few the strongest ones, to produce  the graph as in Fig. \ref{VF_ex1ab}(b), two-dimensional presentation of which is shown in Fig. \ref{VF_ex1ab}(c). Vertices are ordered so that the values of the Fiedler eigenvector, $u_2(n)$, are nondecreasing.  
        	
        	 The signal on this graph  composed of parts of three Laplacian eigenvectors. For the subset of all vertices $\mathcal{V}$, denoted by $\mathcal{V}_1$, which comprises vertices with indices from $m=1$ to $m=40$, the eigenvector with spectral index $k=72$ was used. 
        	For the subset $\mathcal{V}_2$,  with vertex indices from $m=41$ to $m=70$, the signal was equal to the eigenvector $u_k(n)$ with $k=50$. The remaining vertices form the vertex subset $\mathcal{V}_3$, and the signal on this subset was equal to the eigenvector with the spectral index $k=6$. Amplitudes of the eigenvectors were scaled too.
        	
        	 Consider now the  
                vertex-frequency localization kernels, $\mathcal{H}_{m,k}(n)=h_m(n)u_k(n)$, shown in Fig. \ref{VF_windows_LVS}. The constant eigenvector, $u_1(n)=1/\sqrt{N}$, is used in the panel shown in Fig. \ref{VF_windows_LVS}(a) at $m=28$. In this case, the localization window, $h_{28}(n)$, is presented since $\mathcal{H}_{28,1}(n)=h_{28}(n)/\sqrt{N}$. The illustration is repeated in the panel in Fig. \ref{VF_windows_LVS}(c) for the vertex  $m=94$. A frequency shifted version of these two kernels, shown in Figs. \ref{VF_windows_LVS}(a) and (c), are given respectively in Figs. \ref{VF_windows_LVS}(b) and (d), where $\mathcal{H}_{m,21}(n)=h_m(n)u_{21}(n)$ is shown for $m=28$ and $m=94$, respectively.

        The vertex-frequency representation, $S(n,k)$, using the LGFT and the localization window defined in the spectral domain is shown in Fig. \ref{VF_ex1g}. From this representation, we can clearly follow the three constituent signal components, within their intervals of support. The marginal properties, such as the projections of $S(n,k)$ to the vertex index and spectral index axis, are also clearly distinguishable. From the marginal properties we can conclude that the graph signal in hand is spread over all vertex indices, while its spectral localization is dominated by the three spectral indices which correspond to the three components of the original graph signal. In an ideal case of the vertex-frequency analysis, these marginals should  respectively  be equal to $|x(n)|^2$ and $|X(k)|^2$, which is not the case here.

\begin{figure*}
        \centering
        \includegraphics[]{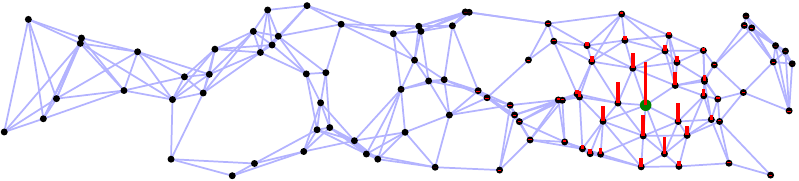}(a)
        \hspace{2mm}
        \includegraphics[]{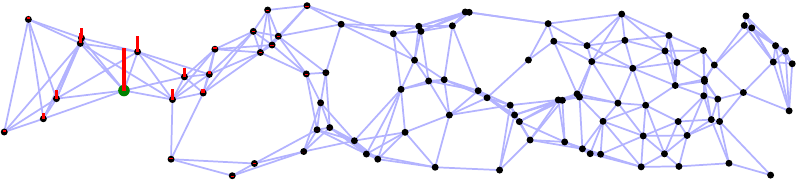}(b)
        
        $\mathcal{H}_{28,1}(n)=h_{28}(n)u_{1}(n) \sim h_{28}(n)$ \hspace{35mm} 
        $\mathcal{H}_{94,1}(n)=h_{94}(n)u_{1}(n) \sim h_{94}(n)$
        
        \vspace{8mm}
        
        \includegraphics[]{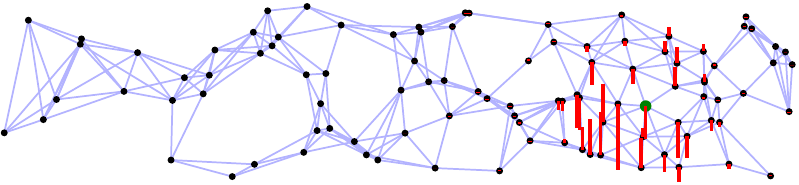}(c)
        \hspace{2mm}
        \includegraphics[]{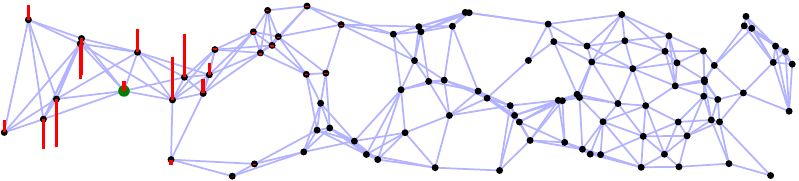}(d)
        
                $\mathcal{H}_{28,21}(n)=h_{28}(n)u_{21}(n)$ \hspace{55mm} 
        $\mathcal{H}_{94,21}(n)=h_{94}(n)u_{21}(n)$
        
        \vspace{4mm}
        
                \includegraphics[scale=0.9]{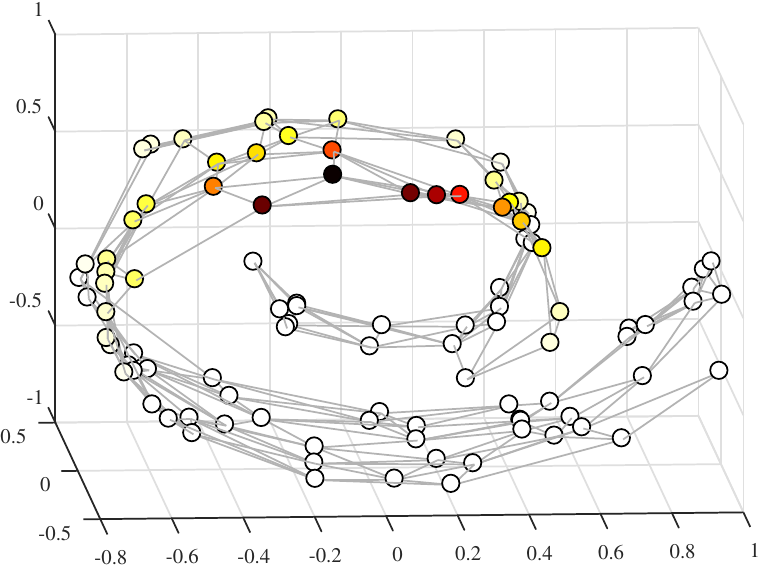} \hspace{6mm} (e)
                \hspace{6mm}
                \includegraphics[scale=0.9]{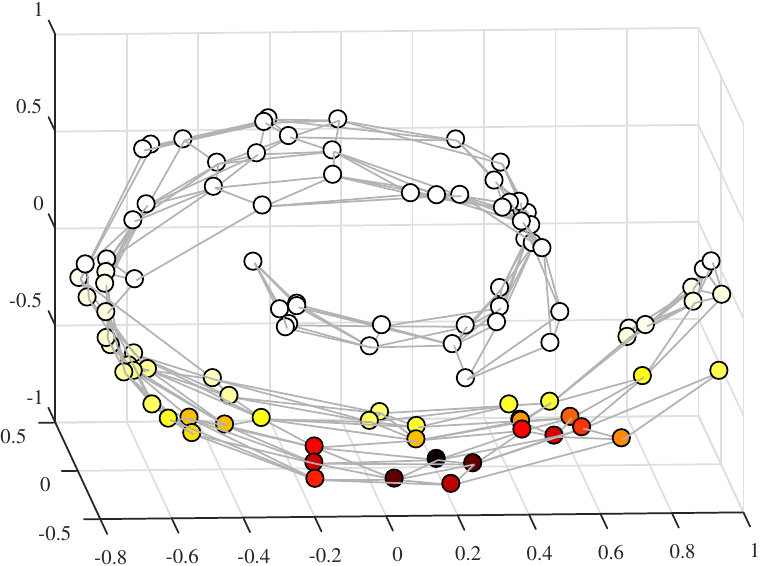} \hspace{8mm} (f)

        $\mathcal{H}_{35,1}(n)=h_{35}(n)u_{1}(n) \sim h_{35}(n)$ \hspace{35mm} 
        $\mathcal{H}_{79,1}(n)=h_{79}(n)u_{1}(n) \sim h_{79}(n)$

        \caption{Illustration of localization kernels, $\mathcal{H}_{m,k}(n)=h_m(n) u_{k}(n)$, for vertex-frequency analysis based on \textit{spectral domain defined windows} in the local graph Fourier transform, $S(m,k)=\sum_{n=1}^N x(n)\mathcal{H}_{m,k}(n)$. (a) Localization kernel $\mathcal{H}_{28,1}(n)=h_{28}(n)u_{1}(n) \sim h_{28}(n)$, for a constant eigenvector, $u_1(n)=1/\sqrt{N}$, centered at the vertex $m=28$. (b) The same localization kernel as in (a) but centered at the vertex $m=94$. (c) Localization kernel, $\mathcal{H}_{28,21}(n)=h_{28}(n) u_{21}(n)$, centered at the vertex $m=28$ and frequency shifted by $u_{21}(n)$. Notice that the variations in kernel amplitude indicate the effects of modulation of the localization window $h_m(n)$. (d) The same localization kernel as in (c), but centered at the vertex $m=94$.
        (e) Three-dimensional representation of kernel $\mathcal{H}_{35,1}(n)=h_{35}(n)u_{1}(n)$, (f) Three-dimensional representation of kernel $\mathcal{H}_{79,1}(n)=h_{79}(n)u_{1}(n)$.
        }
        \label{VF_windows_LVS}
\end{figure*}

\begin{figure}
        \centering
        \includegraphics[scale=0.9]{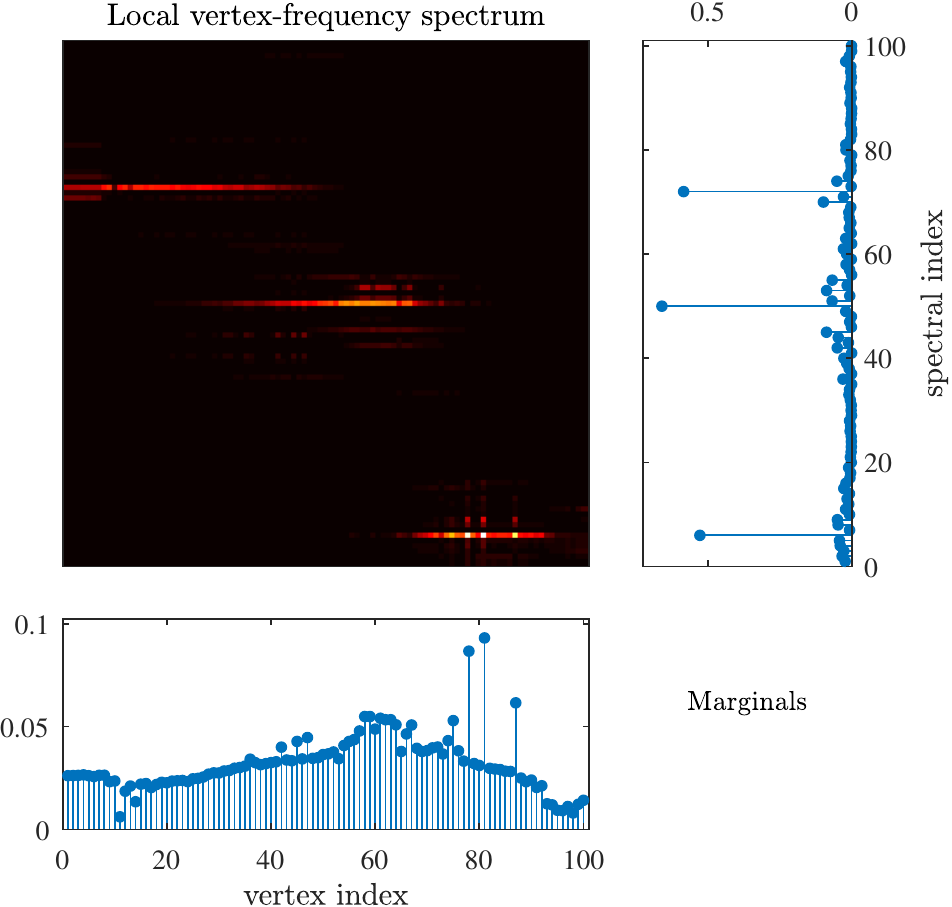}
        \caption{Local vertex-frequency spectrum calculated using the LGFT and the vertex-frequency localized kernels defined in the the spectral domain, as in (\ref{KernelGFDFTmk}). From this representation, we can see that the graph signal consists of three components located at spectral indices $k=72$, $k=50$, and $k=6$, with the corresponding vertex indices subsets $\mathcal{V}_1$, $\mathcal{V}_2$, and $\mathcal{V}_3$, where $\mathcal{V}_1 \cup \mathcal{V}_2 \cup \mathcal{V}_3=\mathcal{V}$. The marginal (vertex and spectrum-wise) properties are shown in the panels right and below the vertex-frequency representation. Observe that, while the graph signal is spread across all vertices, its spectral content is localized at three spectral indices which correspond to the constituent signal components. In an ideal case of vertex-frequency analysis, these marginals should  be respectively equal to $|x(n)|^2$ and $|X(k)|^2$.   }
        \label{VF_ex1g}
\end{figure}

\end{Example}

\subsection{Spectral Domain Localization of the LGFT}

Recall that the classical STFT admits frequency localization using a window in the spectral domain, whereby the dual form of STFT is obtained using the DFT of the original signal and spectral domain window. For graph signals, we can also use this approach to perform localization in the spectral domain, whereby the LGFT is obtained as an inverse GFT of $X(p)$ localized by a spectral domain window, $H(k-p)$, which is centered around spectral index $k$, that is   
\begin{equation}
S(m,k)=\sum_{p=1}^N X(p)H(k-p)\; u_{p}(m). \label{FDLGFT}
\end{equation}
Note that this form of the LGFT can be entirely implemented in the graph spectral domain.  

\begin{Remark}
	The classical time-frequency analysis counterpart of (\ref{FDLGFT}) is \cite{stankovic2014time} $$S(m,k)=\frac{1}{\sqrt{N}}\sum_{p=1}^{N} X(p)H(k-p)e^{j \frac{2 \pi}{N} (m-1)(p-1)},$$  
	where $H(k)$ is a frequency domain window.
\end{Remark}

Note that this form of the LGFT can be entirely implemented in the graph spectral domain. The spectral domain LGFT form in (\ref{FDLGFT}) can be implemented using band-pass transfer functions,  $H_k(\lambda_p)=H(k-p)$, as
\begin{equation}
S(m,k)=\sum_{p=1}^{N} X(p)H_k(\lambda_p)\; u_{p}(m). \label{FDLGFTLam} 
\end{equation}

\begin{Remark}
		The kernel in (\ref{KernelGFDFTmk}) is defined based on low-pass transfer function $H(k)$, which is appropriately shifted in the the spectral domain using the modulation term $u_k(n)$. 		
		The transfer function in (\ref{FDLGFTLam}), $H_k(\lambda_p)$, is  centered (shifted) at a spectral index, $k$, by definition. Hence, in this case, the modulation term $u_k(n)$ is not needed. The kernel is now of the form
		\begin{equation}
		\mathcal{H}_{m,k}(n)=\sum_{p=1}^{N}H_k(\lambda_p)u_p(m)u_p(n). \label{spectral_shifted_kernel}
		\end{equation}	
			\end{Remark}

\subsection{LGFT Realization with Band-Pass Functions}
Assume that the GFT of the localization window, $h_m(n)$, corresponds to a transfer function of a band-pass graph system, centered at an eigenvalue, $\lambda_k$, and around it, and that it is defined in the form of a polynomial given by
\begin{equation}
H_k(\lambda_p)=h_{0,k}+h_{1,k}\lambda_p+\dots+h_{M-1,k}\lambda_p^{M-1}, \label{Huge_LConv}
\end{equation}
{with $M$ being the polynomial order, and $k=0,1,\dots,K-1$, where $K$ is the number of bands.} 
%The vertex shifted version of the window, $h_m(n)$, according to (\ref{GSHVER}), has the GFT in the form  $H_k(\lambda_p)u_p(m)$.  The inverse GFT of this function is the vertex domain kernel centered around a frequency index $k$ and a vertex $m$, that is 

The vertex-frequency transform, $S(m,k)$, for the vertex $m$ and spectral index $k$ then assumes the form as in (\ref{spectral_shifted_kernel}), which can  be written in a vector form as
\begin{align}
\mathbf{s}_k&=\mathbf{U}H_k(\mathbf{\Lambda}) \mathbf{U}^T  \mathbf{x}=H_k(\mathbf{L}) \mathbf{x}=\sum_{p=0}^{M-1}h_{p,k} \mathbf{L}^p\, \mathbf{x}, \label{PolyVert}
\end{align}
where $\mathbf{s}_k$ is the  column vector with elements $S(m,k)$, $m=1,2,\dots,N$, and the property of eigendecomposition of a matrix polynomial is used in this derivation. \textit{In this case, the number of shifted windows, $K$, is not related to the total number of indices $N$. }

\begin{Example}\label{EX:simple_dec}  Consider the simplest decomposition into a low-pass and high-pass part of a graph signal with $K=2$. In this case, two values $k=0$ and $k=1$ represent the low-pass part and high-pass part of the graph signal. Such a decomposition can be achieved by using the graph Laplacian with $h_{0,0}=1$, $h_{0,1}=-1/\lambda_{\max}$, and $h_{1,0}=0$, $h_{1,1}=1/\lambda_{\max}$, where the coefficients are chosen  to form a simple linearly decreasing function of $\lambda_p$ for the low-pass, and a linearly increasing function of $\lambda_p$ for the high-pass, in the corresponding transfer functions. These transfer functions are given by
	\begin{align}
	H_0(\lambda_p)=(1-\frac{\lambda_p}{\lambda_{\max}}),  \,\,\, \,\,\,
	H_1(\lambda_p)=\frac{\lambda_p}{\lambda_{\max}}, \notag
	\end{align}
	further leading to the vertex domain implementation of the LGFT as
	\begin{align}
	\mathbf{s}_0=(\mathbf{I}-\frac{1}{\lambda_{\max}}\mathbf{L})\, \mathbf{x},  \,\,\, \,\,\,
	\mathbf{s}_1=\frac{1}{\lambda_{\max}}\mathbf{L}\, \mathbf{x}. \notag
	\end{align}
	
	To improve the spectral resolution, we can continue with the same transfer function by dividing the low-pass part into its low-pass and high-pass part. The same can be done for the high-pass part, to obtain $$\mathbf{s}_{00}=\Big(\mathbf{I}-\frac{\mathbf{L}}{\lambda_{\max}}\Big)^2 \mathbf{x}, \,\,\,\, \mathbf{s}_{01}=2\Big(\mathbf{I}-\frac{\mathbf{L}}{\lambda_{\max}}\Big)\frac{\mathbf{L}}{\lambda_{\max}} \mathbf{x}, \,\,\,\, \mathbf{s}_{11}=\frac{\mathbf{L}^2}{\lambda^2_{\max}} \mathbf{x}.$$  
	the factor of 2 appears in the new middle pass-band, $\mathbf{s}_{01}$, since the low-high-pass and the high-low-pass components are the same. 
	
	The division can be done into $K$ bands corresponding to the terms of a binomial form $$\Big((\mathbf{I}-\mathbf{L}/\lambda_{\max})+\mathbf{L}/\lambda_{\max}\Big)^K \, \mathbf{x},$$ with the transfer functions in the vertex domain given by
	$$H_k(\mathbf{L})={K \choose k}\Big(\mathbf{I}-\frac{1}{\lambda_{\max}}\mathbf{L}\Big)^{K-k}\Big(\frac{1}{\lambda_{\max}}\mathbf{L}\Big)^k.$$
\end{Example}
\begin{Example}\label{EX:Tharf} The transfer functions $H_k(\lambda_p)$, $p=1,2,\dots,N$, $k=0,1,\dots,K-1$ in the spectral domain corresponding to the binomial form terms for $K=26$ are shown in Fig. \ref{test_bands_a}(a). These functions are used for the LGFT calculation at vertex indices $m=1,2,\dots,N$ in $k=0,1,\dots,K-1$ bands. Since the bands are quite spread, the resulting LGFT is spread along the frequency axis. The concentration can be improved by reassigning the values of $S(m,k)$ to the position of their maximum value along frequency band index, $k$, for each vertex index, $m$. Such a reassigned LGFT values are given in Fig. \ref{test_bands_b} (a).  
	
	\begin{figure}
		\centering
		\includegraphics[]{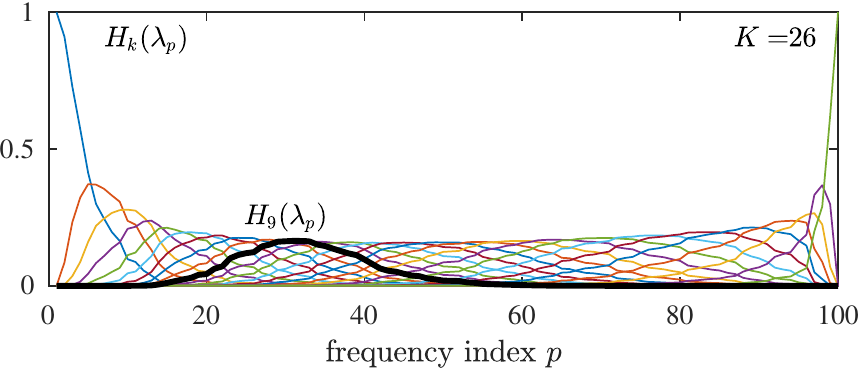}(a)
		
		\vspace{1.5mm}
		
		\includegraphics[]{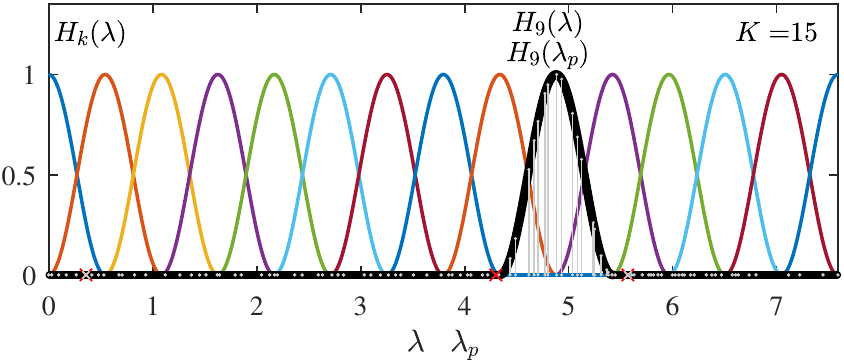}(b)
		
		\vspace{1.5mm}
		
		\includegraphics[]{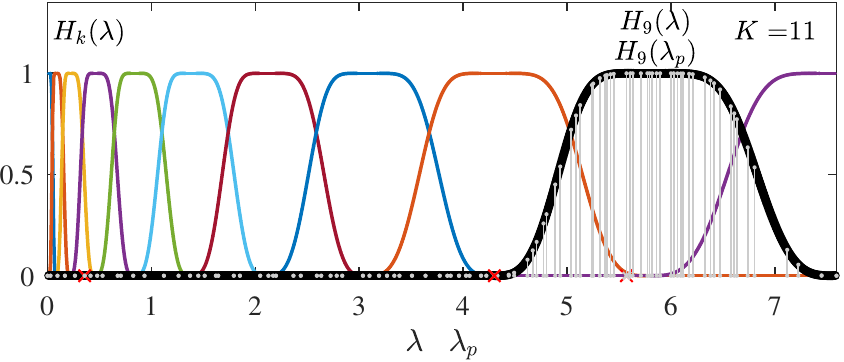}(c)
		
			\vspace{1.5mm}
		
		\includegraphics[]{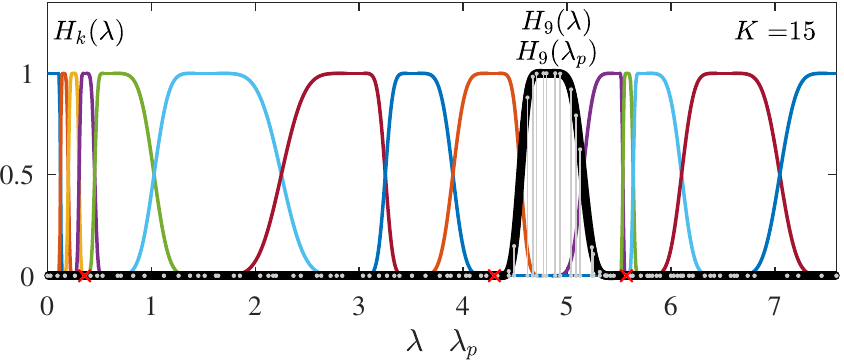}(d)
		
			\vspace{1.5mm}
		
			\includegraphics[]{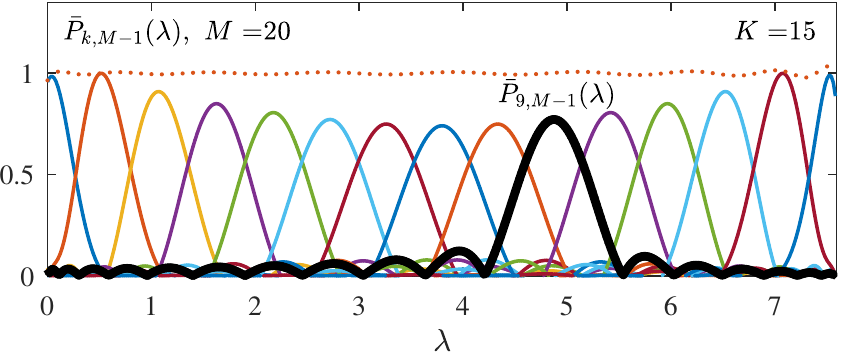}(e)

		\caption{Exemplar transfer functions in the spectral domain. (a) The spectral domain transfer functions $H_k(\lambda_p)$, $p=1,2,\dots,N$, $k=0,1,\dots,K-1$ which correspond to the binomial form terms for $K=26$. (b) The transfer functions $H_k(\lambda_p)$, $p=1,2,\dots,N$, $k=0,1,\dots,K-1$ which correspond to the half-cosine form terms for $K=15$. (c) The spectral index-varying (wavelet-like) transfer functions $H_k(\lambda_p)$, $p=1,2,\dots,N$, $k=0,1,\dots,K-1$ which correspond to the half-cosine form terms for $K=11$. (d) The spectral domain signal adaptive transfer functions $H_k(\lambda_p)$, $p=1,2,\dots,N$, $k=0,1,\dots,K-1$ which satisfy the condition $\sum_{k=0}^{K-1}H_k^2(\lambda_p)=1$, with $K=17$. (e)    Polynomial Chebyshev approximations of transfer functions from panel (b), $\bar{P}_{k,M-1}(\lambda),k=0,1,\dots,K-1$, with $M=20$, where  dotted horizontal line  designates $\sum_{k=0}^{K-1}H_k(\lambda)$.
				The transfer function $H_9(\lambda)$ is designated by the thick black line for each considered domain in panels (a) -- (d), while its discrete values at $\lambda_p$, $H_9(\lambda_p)$, are shown in gray in panels (b) -- (d).  
		} 
		\label{test_bands_a}
	\end{figure}
	
	\begin{figure*}
		\centering
		
		 \hspace{1.26mm}\includegraphics[scale=0.7]{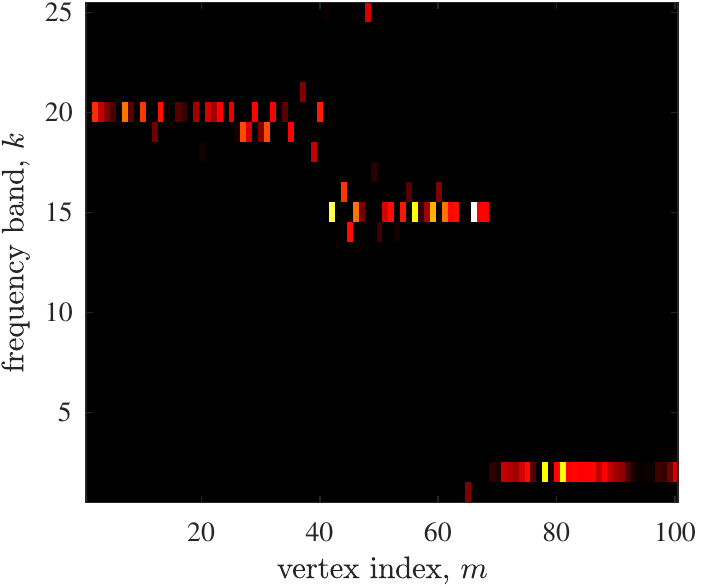}\hspace{3mm}(a) 
				\includegraphics[scale=0.7]{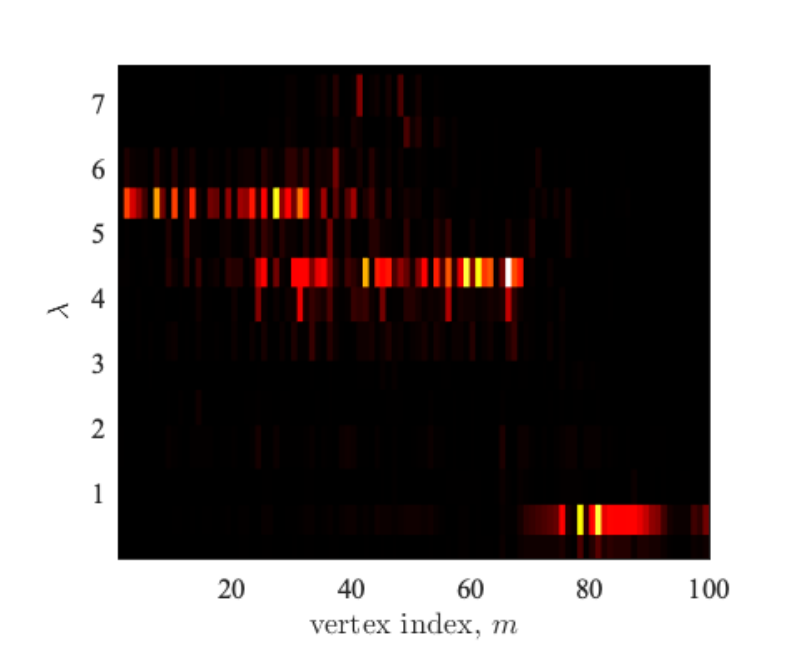}(b)
					%\hspace{2mm}
				\includegraphics[scale=0.7]{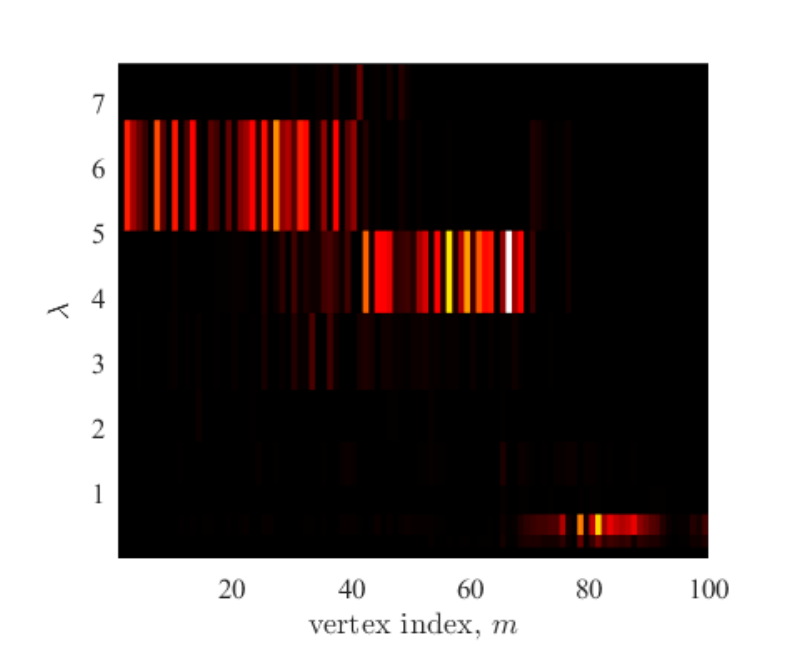}(c) 				
		\vspace{2mm}
		
		\includegraphics[scale=0.7]{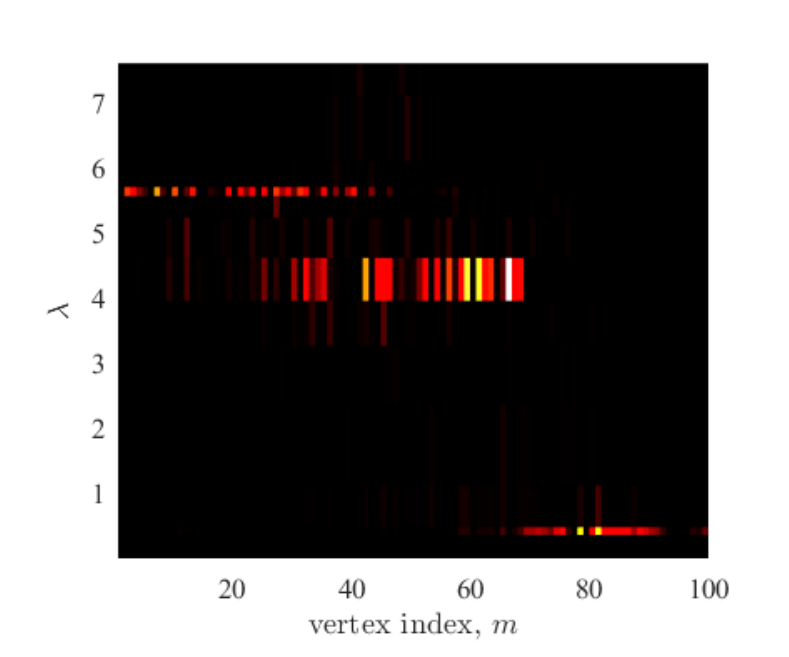}(d)
		 	\hspace{-2mm}
			\includegraphics[scale=0.7]{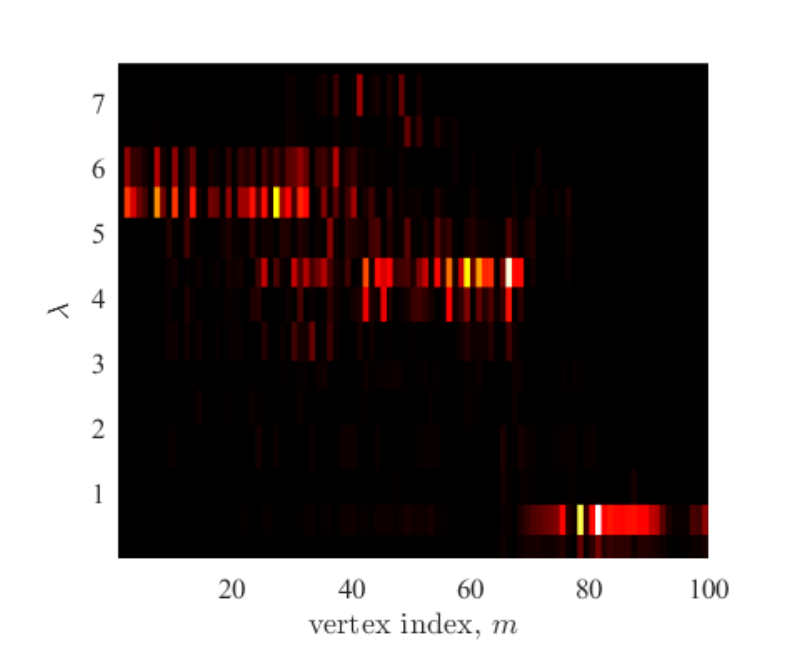}(e)
			%	\hspace{2mm}
		\includegraphics[scale=0.7]{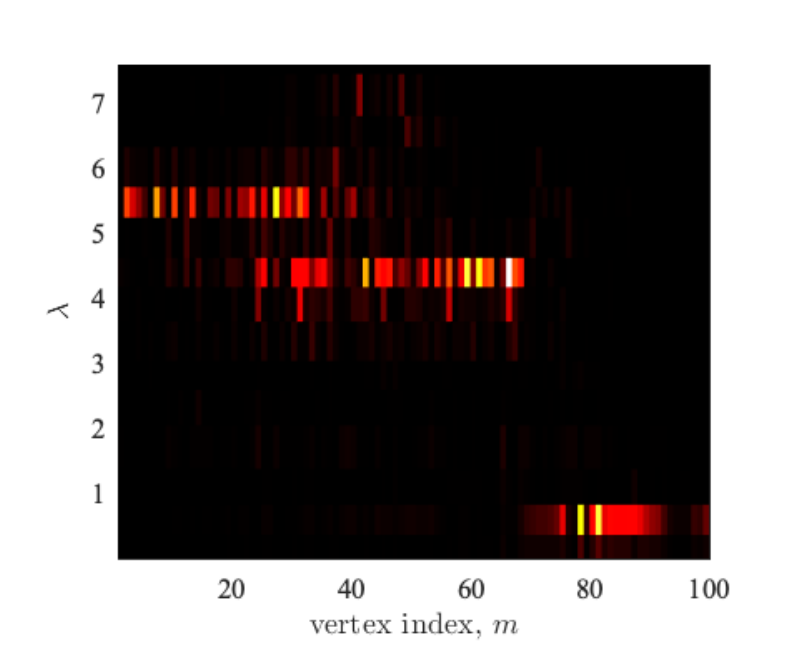}(f)
		
		\caption{Vertex-frequency representation of a three-component signal in Fig. \ref{VF_ex1ab}(d). (a) The LGFT of the signal is calculated by using the transfer functions in Fig. \ref{test_bands_a}(a) for frequency selection.(b) The LGFT of the signal from Fig \ref{VF_ex1ab}(d), calculated by using the transfer functions in Fig. \ref{test_bands_a}(b) for frequency selection. The LGFT values, $S(m,k)$, were reassigned to the position of its maximum value along the frequency band index, $k$, for each vertex index, $m$. (c) The LGFT of the signal from Fig \ref{VF_ex1ab}(d), calculated using the wavelet-like transfer functions in Fig. \ref{test_bands_a}(c) for frequency selection.   (d) The LGFT of signal from Fig \ref{VF_ex1ab}(d), using signal adaptive transfer functions from Fig. \ref{test_bands_a}(c) for frequency selection.   (e) The LGFT of signal from Fig \ref{VF_ex1ab}(d), calculated based on the Chebyshev approximation of band-pass transfer functions from Fig. \ref{test_bands_a} (b) with $M=20$, shown in \ref{test_bands_a} (e). (f) The LGFT of signal from Fig \ref{VF_ex1ab}(d), calculated based on the Chebyshev approximation of band-pass transfer functions from Fig. \ref{test_bands_a} (b) with $M=50$. LGFT from (e) and (f) illustrate the influence of the Chebyshev approximation convergence on the final vertex-frequency representation. In all considered cases, the LGFT values, $S(m,k)$, were reassigned to the position of its maximum value along the frequency band index, $k$, for each vertex index, $m$. }
		\label{test_bands_b}
	\end{figure*}
	
	%\begin{figure}
	%       \centering
	%       \includegraphics[]{test_bands_c}
	%       \caption{XXX bands c}
	%       \label{test_bands_c}
	%\end{figure}  
	
\end{Example}    

Of course, any other set of band-pass functions, $H_k(\mathbf{\Lambda})$, can be used to produce the LGFT from (\ref{FDLGFTLam}) in the form (\ref{PolyVert}), as
\begin{align}
\mathbf{s}_k&=H_k(\mathbf{L}) \mathbf{x}. \label{GenerFunVert}
\end{align}
Commonly used examples of such band-pass functions are the spline or raised cosine (Hann window) functions. We will next use the general form of the shifted raised cosine functions as the transfer functions, defined by  
\begin{equation}
H_k(\lambda)\!\!=\!\!\begin{cases}
\!\sin^2\!\bigg(\frac{\pi}{2}\frac{a_k}{b_k-a_k}(\frac{\lambda}{a_k}-1)\bigg), \text{ for } a_k<\!\lambda \le \! b_k \\
\!\cos^2\!\bigg(\frac{\pi}{2}\frac{b_k}{c_k-bk}(\frac{\lambda}{b_k}-1)\bigg), \text{ for } b_k<\!\lambda \le \! c_k \\
0 , \text{ elsewhere, }
\end{cases} 
\label{defHannWGen}
\end{equation} 
where $(a_k,b_k]$ and $(b_k,c_k]$, $k=0,1,\dots,K-1$, define the spectral bands for $H_k(\mathbf{\Lambda})$. For uniform bands within $0\le \lambda \le \lambda_{\max}$, the intervals can be defined by 
\begin{gather}
a_k=a_{k-1}\!+\!\frac{\lambda_{\max}}{K\!\!-\!1}, \,
b_k=a_k\!+\!\frac{\lambda_{\max}}{K\!\!-\!1}, \,
c_k=a_k\!+\!2\frac{\lambda_{\max}}{K\!\!-\!1} \label{intervalsabc}
\end{gather} 
with $a_1=0$ and $\lim_{\lambda\to 0}(a_1/\lambda)=1$. 
The initial transfer function, $H_0(\lambda)$, is defined using only $0=b_0\le \lambda \le c_0=\lambda_{\mathrm{max}}/K$, while the last transfer function, $H_{K-1}(\lambda)$, is defined using the interval $a_{K-1} < \lambda \le b_{K-1}=\lambda_{\max}$  in (\ref{defHannWGen}).  

The raised cosine transfer function in (\ref{defHannWGen}) satisfy the following condition
\begin{equation}
\sum_{k=0}^{K-1}H_k(\lambda_p)=1. \label{sumCondition}
\end{equation}

If the squares are omitted in (\ref{defHannWGen}), the condition 
\begin{equation}
\sum_{k=0}^{K-1}H_k^2(\lambda_p)=1 \label{squaredsumCondition}
\end{equation}
would be satisfied. 

In this case, the transfer functions become \cite{leonardi2013tight}
 \begin{equation}
H_k(\lambda)=\begin{cases}
\sin\bigg(\frac{\pi}{2}v_x\Big(\frac{a_k}{b_k-a_k}(\frac{\lambda}{a_k}-1)\Big)\bigg), \text{ for } a_k<\lambda \le b_k \\
\cos\bigg(\frac{\pi}{2}v_x\Big( \frac{b_k}{c_k-bk}(\frac{\lambda}{b_k}-1)\Big)\bigg), \text{ for } b_k<\lambda \le c_k \\
0 , \text{ elsewhere, }
\end{cases} 
\label{defMeyeWGen}
\end{equation}  
with $a_{k+1}=b_k$, $b_{k+1}=c_k$ and the initial and the last intervals defined as in (\ref{intervalsabc}). Since the sine and cosine functions are not differentiable at the interval-end points, the argument, $x=\frac{a_k}{b_k-a_k}(\frac{\lambda}{a_k}-1)$, is commonly mapped as  $v_x(x)=x^4(35-84x+70x^2-20x^3)$, for $0\le x \le 1$, producing Meyer's wavelet-like transfer functions. 

The conditions for graph signal reconstruction from the LGFT will be discussed  in Section \ref{GFTInversion} and both forms, (\ref{sumCondition}) and (\ref{squaredsumCondition}), will be used.

\begin{Example} The experiment from Example \ref{EX:Tharf} is repeated with the raised cosine transfer functions $H_k(\lambda_p)$, $p=1,2,\dots,N$, $k=0,1,\dots,K-1$, shown in Fig. \ref{test_bands_a}(b) for $K=15$. These functions are used for the LGFT calculation at vertex indices $m=1,2,\dots,N$ in $k=0,1,\dots,K-1$ bands. The  LGFT values, appropriately reassigned to each eigenvalue, $\lambda_p$, are given in Fig. \ref{test_bands_b}(b). 	

	Moreover, the experiment is extended by considering a spectral index-varying (wavelet-like transform) form of Meyer's transfer functions, $H_k(\lambda_p)$, $p=1,2,\dots,N$, $k=0,1,\dots,K-1$, as depicted in Fig. \ref{test_bands_a}(c).  The so-obtained LGFT values are shown in Fig. \ref{test_bands_b}(c). In order to illustrate the change of resolution in this case, the LGFT were reassigned to each corresponding eigenvalue, $\lambda_p$.  
		Notice that the transfer functions, $H_k(\lambda_p)$, in this example, satisfy the condition in (\ref{squaredsumCondition}) 	which will be important for the frame-based LGFT inversion.   
\end{Example}

The spectral graph wavelet-like transform is just an example  of  varying  spectral transfer functions in the LGFT, with the highest spectral resolution (the narrowest spectral wavelet functions) for small values of the smoothness index, $\lambda_p$. The spectral resolution decreases as the spectral wavelet functions become wider for large smoothness index values, Fig. \ref{test_bands_a}(c). In general, the change of resolution may be arbitrary and signal adaptive, for example, the resolution may be higher for the spectral intervals of $\lambda$ which are rich in signal components and lower within the intervals where there are no signal components. 

\begin{Example} The concept of signal adaptive intervals for transfer functions of the form (\ref{defMeyeWGen}) is illustrated in Fig. \ref{test_bands_a}(d), where small intervals are defined around a significant spectral content, while wide intervals are used for $\lambda$ corresponding to the less significant signal spectral content \cite{behjat2016signal,behjat2019spectral}. In the considered case, narrowest window intervals (corresponding to the high resolution in the vertex-frequency representation) are defined around the spectral positions of three signal components, at  $\lambda=0.36$, $\lambda=4.30$, and $\lambda=5.57$. Vertex-frequency representation calculated based on these transfer functions is shown in Fig. \ref{test_bands_b}(d), and the assigned eigenvalue as a spectral axis. A notably high resolution in the representation of the first and the third components at $\lambda=0.36$ and $\lambda=5.57$ is the result of the fine interval definition for spectral windows in Fig.  \ref{test_bands_a}(d). This particular choice of the intervals  allowed for high spectral resolution representation with a small number of transfer functions, $K=15$. A wider interval width for the third component resulted in a lower spectral resolution than in the case of the other two components.

	%The transfer functions of the form defined in (\ref{MeyerdefHannWGen}) are used with signal adaptive intervals. These intervals are defined in such a way that they are small (fine) around $\lambda$, where a significant signal spectral content is detected, and are big (rough) around $\lambda$  where the signal spectral content is low, as in Fig. \ref{test_bands_bCSA}(a) and (b). The intervals are narrow (with a high resolution) around the three signal components at $\lambda=0.38$, $\lambda=1.87$, and $\lambda=4.62$.
	
	% Vertex-frequency representation with these transfer functions is shown in Fig. \ref{test_bands_bCSA}(c) and (d) with the spectral band index, $k$, and the assigned eigenvalue (spectral) index, $p$, as a spectral axis. Fine intervals around the spectral signal components allowed for high spectral resolution representation, as in  Fig. \ref{test_bands_bCSA}(c), with a smaller number of transfer functions $K+1=17$. 
\end{Example}

\subsection{Polynomial LGFT Approximation}\label{PolyAppo}
Bandpass LGFT functions, $H_k(\lambda),~k=0,1,\dots,K-1$, of the form (\ref{defHannWGen}) or (\ref{defMeyeWGen}) can be implemented using a polynomial approximation of order $(M-1)$,  $\bar{P}_{k,M-1}(\lambda)$, $k=0,1,\dots,K-1$. Since the transfer functions $H_k(\lambda)$ are used at discrete points $\lambda=\lambda_p$ and the polynomial approximation is continuous within $0\le \lambda \le \lambda_{\max}$, the natural choice are the so called “min-max”  Chebyshev polynomials whose maximal deviation from the desired transfer function is minimal. 

The Chebyshev polynomials are defined by 
\begin{gather*}
T_0(z)=1, \,\,\, T_1(z)=z,\dots, T_m(z)=2zT_{m-1}(z)-T_{m-1}(z).
\end{gather*}
for $m\ge 2$ and $-1\le z \le 1$. 
 
The finite $(M-1)$-order  of the Chebyshev polynomials
\begin{equation}
\bar{P}_{k,M-1}(\lambda)=\frac{c_{k,0}}{2}+\sum_{m=1}^{M-1}c_{k,m}\bar{T}_m(\lambda),
\label{cheb_approx}
\end{equation}
where $\bar{T}_m(\lambda)=T_m(2\lambda/\lambda_{\max}-1)$ in order to map the argument to the interval, $0\le \lambda \le \lambda_{\max}$, to the interval from $-1$ to $1$. The polynomial coefficients are calculated using the Chebyshev polynomial inversion property as 
$c_{k,m}=\frac{2}{\pi}\int_{-1}^{1}H_k((z+1)\lambda_{\max}/2)T_m(z)dz/\sqrt{1-z^2}$.

This leads to the vertex domain implementation (\ref{PolyVert}) of the spectral LGFT form, given by
$$\mathbf{s}_k=\bar{P}_{k,M-1}(\mathbf{L})\mathbf{x},$$
for $k=0,1,2,\dots,K-1,$ with
\begin{gather}
\bar{P}_{k,M-1}(\mathbf{L})=\frac{c_{k,0}}{2}+\sum_{m=1}^{M-1}c_{k,m}\bar{T}_m(\mathbf{L}), \label{plylgft} \\
=h_{0,k}\mathbf{I}+h_{1,k}\mathbf{L}+h_{2,k}\mathbf{L}^2+\dots+h_{(M-1),k}\mathbf{L}^{M-1}. \nonumber
\label{vd_implementation}
\end{gather}
The polynomial form in (\ref{plylgft}) uses only  the
$(M-1)$-neighborhood in calculation of the LGFT for each considered vertex, without the need for eigendecomposition analysis, thus significantly reducing the computational cost.

\begin{Example}\label{ExLGFTCHP}
	Consider the shifted transfer functions, $H_k(\lambda),~k=0,1,\dots,K-1$, defined by  (\ref{defHannWGen}) and (\ref{intervalsabc}), shown in Fig. \ref{test_bands_a}(b), for $K=15$. Functions $H_k(\lambda)$ satisfy $\sum_{k=0}^{K-1}H_k(\lambda)=1$.
	 Each individual transfer function,  $H_k(\lambda)$, is approximated using the Chebyshev polynomial, $\bar{P}_{k,M-1},k=0,1,\dots,K-1$, of the form (\ref{cheb_approx}). Two orders of the polynomial are considered: $M=20$ and $M=40$. The Chebyshev polynomial approximations of the band-pass functions from Fig. \ref{test_bands_a}(b) are shown in Fig. \ref{test_bands_a}(e), for $M=20$. For convenience, summation $\sum_{k=0}^{K-1}\bar{P}_{k,M-1}(\lambda)$ is also calculated. As designated by the horizontal dotted line in Fig. \ref{test_bands_a}(e), summation values are close to 1, thus indicating the stable invertibility of the LGFT.
	 
	\begin{table}
		\centering
		\caption{Coefficients, $h_{i,k}$, $i=0,1,\dots,M-1$, $k=0,1,\dots,K-1$,  for the polynomial calculation of the LGFT, $\mathbf{s}_k$, of a signal, $\mathbf{x}$ , in various spectral bands, $k$, for $(M-1)=5$ and $K=10$. \vspace*{2mm}}
		\small
		\setlength{\tabcolsep}{3pt}
		
		\begin{tabular}{crrrrrr}
			\multicolumn{7}{c}{$\mathbf{s}_k=(h_{0,k}\mathbf{I}+h_{1,k}\mathbf{L}+h_{2,k}\mathbf{L}^2+h_{3,k}\mathbf{L}^3+h_{4,k}\mathbf{L}^4+h_{5,k}\mathbf{L}^5) \mathbf{x}$} \\
			\midrule
			$k$ & $h_{0,k}$ & $h_{1,k}$ & $h_{2,k}$ & $h_{3,k}$ & $h_{4,k}$ & $h_{5,k}$ \\
			\midrule
		
	  0 & $ 1.079$ & $-1.867$ & $ 1.101$ & $-0.2885$ & $ 0.03458$ & $-0.001548$ \\
	1 & $-0.053$ & $ 1.983$ & $-1.798$ & $ 0.5744$ & $-0.07722$ & $ 0.003723$ \\
	2 & $-0.134$ & $ 0.763$ & $-0.310$ & $ 0.0222$ & $ 0.00422$ & $-0.000460$ \\
	3 & $ 0.050$ & $-0.608$ & $ 0.900$ & $-0.3551$ & $ 0.05348$ & $-0.002762$ \\
	4 & $ 0.096$ & $-0.726$ & $ 0.768$ & $-0.2475$ & $ 0.03172$ & $-0.001424$ \\
	5 & $ 0.016$ & $-0.013$ & $-0.128$ & $ 0.1047$ & $-0.02231$ & $ 0.001424$ \\
	6 & $-0.073$ & $ 0.616$ & $-0.779$ & $ 0.3228$ & $-0.05135$ & $ 0.002762$ \\
	7 & $-0.051$ & $ 0.351$ & $-0.356$ & $ 0.1146$ & $-0.01323$ & $ 0.000460$ \\
	8 & $ 0.084$ & $-0.687$ & $ 0.871$ & $-0.3751$ & $ 0.06409$ & $-0.003723$ \\
	9 & $-0.021$ & $ 0.183$ & $-0.251$ & $ 0.1172$ & $-0.02196$ & $ 0.001419$ \\
			\bottomrule
		\end{tabular}
		\label{tablecoeffLGHT}
	\end{table}
	
	The approximations of transfer functions, $H_k(\lambda)$, obtained in this way, are used for the  LGFT based vertex-frequency analysis of the three-component signal from Fig. \ref{VF_ex1ab}(d). The absolute LGFT values are presented in Fig. \ref{test_bands_b}(e) for $M=20$ and in Fig. \ref{test_bands_b}(f) for $M=40$ (almost the same as in Fig. \ref{test_bands_b}(b)). Lower component concentration in Fig. \ref{test_bands_b}(d) than in Fig. \ref{test_bands_b}(e) is related to the less precise approximation of the spectral transfer functions for $M=20$ than for $M=40$. Notice that high values of the polynomial order, $(M-1)$, increase calculation complexity and require wide vertex neighborhood in the calculation of the LGFT.
	
\end{Example}
\begin{Example}
	Chebyshev polynomial approximation of order $(M-1)=5$ is calculated for the band-pass  transfer functions, $H_k(\lambda),~k=0,1,\dots,K-1$, of the raised cosine form (\ref{defHannWGen}), with $K=10$. The corresponding coefficients, $h_{i,k}$  for the vertex-domain implementation (\ref{PolyVert}),  are given in  Table \ref{tablecoeffLGHT}.
\end{Example}

\subsection{The LGFT and the Graph Wavelet Transform }

As in classical signal processing, wavelet coefficients can be defined as a \textit{projection of a graph signal onto the wavelet kernel functions}. Assume that the basic form for the wavelet definition in the spectral domain is $H(\lambda_p)$. The wavelet in spectral domain then represents a scaled version of $H(\lambda_p)$ in scale $s_i$, $i=1,2,\dots,K-1$, and is denoted by $H_{s_i}(\lambda_p)=H(s_i\lambda_p)$ \cite{hammond2019spectral,behjat2019spectral,Behjat2015,rustamov2013wavelets,jestrovic2017fast,masoumi2019shape}. Additionally,  a low-pass scale (father wavelet) function $G(\lambda_p)$, plays the role of low-pass function, $H_0(\lambda_p)$, in the LGFT. Therefore, a set of discrete scales for the wavelet calculation, denoted by $s \in \{s_1, s_2,\dots,s_{K-1}\}$, is assumed with corresponding spectral transfer functions, $H_{s_i}(\lambda_p)$ and  $G(\lambda_p)$. Now, in the same way as in the case of the kernel form of the LGFT in (\ref{KernelGFDFTDEF}), 
 the graph wavelet transform is defined using the band-pass scaled wavelet kernel, $\psi_{m,{s_i}}(n)$, instead of the LGFT kernel, $\mathcal{H}_{m,k}(n)$, in (\ref{spectral_shifted_kernel}). This yields
\begin{equation}
\psi_{m,{s_i}}(n)= \sum_{p=1}^NH(s_i\lambda_p)u_p(m)u_p(n), \label{Waveletmk}
\end{equation}
with  the wavelet coefficients given by
\begin{gather*}
W(m,s_i)= \sum_{n=1}^N \psi_{m,{s_i}}(n)x(n)= \\
 \sum_{n=1}^N \sum_{p=1}^N  H(s_i\lambda_p)x(n)u_p(m)u_p(n)
= \sum_{p=1}^N  H(s_i\lambda_p)X(p)u_p(m).
\end{gather*}
The Meyer wavelet like transfer functions in the spectral domain, $H(s_i\lambda_p)$, are defined  in (\ref{defMeyeWGen}) with the argument $v_x(q(s_i\lambda-1))$ and the following  intervals of the support: $1/s_1<\lambda \le M/s_1$ for $i=1$ (sine function in (\ref{defMeyeWGen})) and $1/s_i <\lambda \leq M/s_i$ (sine functions), $M/s_i <\lambda \leq M^2/s_i$ (cosine functions), for $i=2,3,\dots,K-1$, where $q=1/(M-1)$ and the scales are related as $s_i=s_{i-1}M=M^i/\lambda_{\max}$. The interval for the low-pass function, $G(\lambda)$, is $0 \le \lambda \le M^2/s_{K-1}$ (cosine function within $M/s_{K-1} <\lambda \leq M^2/s_{K-1}$ and the value $G(\lambda)=1$ as  $\lambda \to 0$). 

In the implementations, we can use the vertex domain localized polynomial approximations of the spectral wavelet functions, in the same way as described in Section \ref{PolyAppo}.

\subsection{Windows Defined Using the Vertex Neighborhood}

The window, $h_m(n)$, localized at a vertex $m$ can also be defined using the vertex neighborhood. Recall that the distance between vertices $m$ and $n$, $d_{mn}$, is equal to the length of the shortest walk from vertex $m$ to vertex $n$, and that $d_{mn}$ are integers. Then, the window function can be defined as
$$
h_m(n)=g(d_{mn}),
$$ 
where $g(d)$ corresponds to any basic window function in classical signal processing.  For example, the Hann window can be used, which is defined as
$$
h_m(n)=\frac{1}{2}(1+\cos(\pi d_{mn}/D)), \text{ for } 0\le d_{mn} < D,
$$
where $D$ is the assumed window width. 

For convenience, window functions for every vertex can be calculated in a matrix form as follows:
\begin{itemize}
        \item
 The vertices for which the distance is $d_{mn}=1$ are defined with an adjacency (neighborhood one) matrix $\mathbf{A}_1=\mathbf{A}$. The vertices which belong to the one-neighborhood of a vertex, $m$, are indicated by the unit-value elements in the $m$th row of the adjacency matrix $\mathbf{A}$ (in unweighted graphs).  In weighed  graphs, the corresponding adjacency matrix $\mathbf{A}$ can be obtained from the weighting matrix  $\mathbf{W}$ as $\mathbf{A}=\operatorname{sign}( \mathbf{W})$.
 \item  
 Vertices $m$ and $n$, for which the distance is $d_{mn}=2$ are defined by the following matrix
$$
\mathbf{A}_2=(\mathbf{A} \odot \mathbf{A}_1 ) \circ (\mathbf{1}-\mathbf{A}_1) \circ (\mathbf{1}-\mathbf{I}),
$$
where $\odot$ is the logical (Boolean) matrix product, $\circ$ is the Hadamard (element-by-element) product, and $\mathbf{1}$ is a matrix with all elements equal to 1. The $m$th row of matrix $\mathbf{A} \odot \mathbf{A}_1$ gives information about all vertices that are connected to the vertex $m$ with walks of length $K=2$ or lower.  It should be mentioned that the element-by-element multiplication of  $(\mathbf{A} \odot \mathbf{A}_1 )$ by matrix $(\mathbf{1}-\mathbf{A}_1)$ removes the vertices connected with walks of length $1$, while the multiplication by $(\mathbf{1}-\mathbf{I})$ removes its diagonal elements. 
\item
For $d_{mn}=d\ge 2 $,  we arrive at a recursive relation for the calculation of a matrix which will give the information about the vertices separated by  distance $d$. Such a matrix has the form
\begin{equation}
\mathbf{A}_{d}= (\mathbf{A} \odot \mathbf{A}_{d-1} ) \circ (\mathbf{1}-\mathbf{A}_{d -1})\circ (\mathbf{1}-\mathbf{I}). \label{AdDEF}
\end{equation}
\end{itemize}

The window matrix for an assumed graph window width, $D$, can now be defined as
$$ 
\mathbf{P}_{D}= g(0)\mathbf{I}+g(1)\mathbf{A}_1+\dots+g(D-1)\mathbf{A}_{D-1},  
$$
so that a graph signal, localized around vertex $m$, may be formed based on this matrix, as 
$$
x_m(n)=h_m(n)x(n)=P_D(n,m)x(n).
$$

The LGFT representation of a graph signal, $x(n)$, then becomes
\begin{equation}
S(m,k)=\sum_{n=1}^N x(n)h_m(n)\; u_{k}(n)=\sum_{n=1}^N x(n)P_D(n,m)\; u_{k}(n),
\end{equation}
with the vertex-frequency kernel given by 
\begin{equation}
\mathcal{H}_{m,k}(n)=h_m(n)u_k(n)=P_D(n,m) u_{k}(n). \label{KernelGFDFTmkA}
\end{equation}
This allows us to arrive at the matrix form of the LGFT, given by
\begin{equation}
\mathbf{S}=\mathbf{U}^{T} (\mathbf{P}_D \circ  [\mathbf{x},\  \mathbf{x},\ \ldots ,\ \mathbf{x}]),
\end{equation}
where $[\mathbf{x},\  \mathbf{x},\ \ldots, \ \mathbf{x}]$ is an $N\times N$ matrix  the columns of which are signal vector $\mathbf{x}$.

For a rectangular function $g(d)=1$, and for any $d<D$, the LGFT can be calculated recursively with respect to the window width, $D$, as
\begin{equation}
\mathbf{S}_D = \mathbf{S}_{D-1} + \mathbf{U}^{T} (\mathbf{A}_{D-1} \circ [\mathbf{x},\  \mathbf{x},\ \ldots, \ \mathbf{x}]).
\end{equation}

        \begin{Example} Consider the local vertex-frequency representation of the signal from Fig. \ref{VF_ex1ab}, using vertex domain defined windows. 
        
        The  localization kernels, $\mathcal{H}_{m,k}(n)=h_m(n)u_k(n)$, are shown in Fig. \ref{VF_windows_neigh} for two vertices and two spectral indices. Observe that for the spectral index $k=1$, the localization kernel is proportional to the localization function $h_m(n)$, given in \ref{VF_windows_neigh}(a) and (c) for the vertices  $m=28$ and  $m=94$. Frequency modulated forms of these localization functions are shown in Figs. \ref{VF_windows_neigh}(b) and (d), for the same vertices and $k=21$. 
        
        The vertex domain window is next used to analyze  the graph signal  from Fig. \ref{VF_ex1ab}. Vertex-frequency representation, $S(n,k)$, obtained with the LGFT and the vertex domain  localization window is given in Fig. \ref{VF_ex1h}. Again, we can observe three graph signal components in three vertex regions. The marginals of  $S(n,k)$ are also shown at the right and bottom panels.  

\begin{figure*}
        \centering
        \includegraphics[]{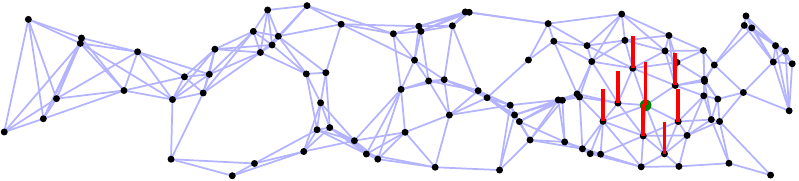}(a)
        \hspace{2mm}
        \includegraphics[]{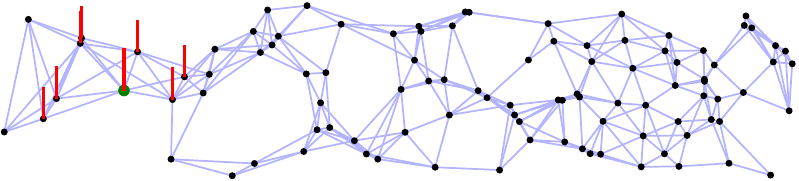}(b)
        
                $\mathcal{H}_{28,1}(n)=h_{28}(n)u_{1}(n) \sim h_{28}(n)$ \hspace{35mm} 
        $\mathcal{H}_{94,1}(n)=h_{94}(n)u_{1}(n) \sim h_{94}(n)$
        
        \vspace{8mm}
        
        \includegraphics[]{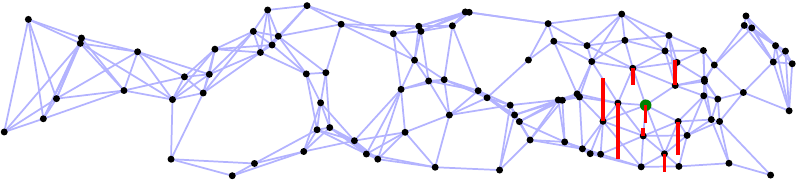}(c)
        \hspace{2mm}
        \includegraphics[]{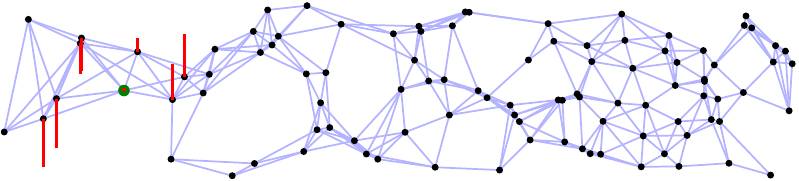}(d)
        
                $\mathcal{H}_{28,21}(n)=h_{28}(n)u_{21}(n)$ \hspace{55mm} 
        $\mathcal{H}_{94,21}(n)=h_{94}(n)u_{21}(n)$
        
         \vspace{4mm}
        
         \includegraphics[scale=0.9]{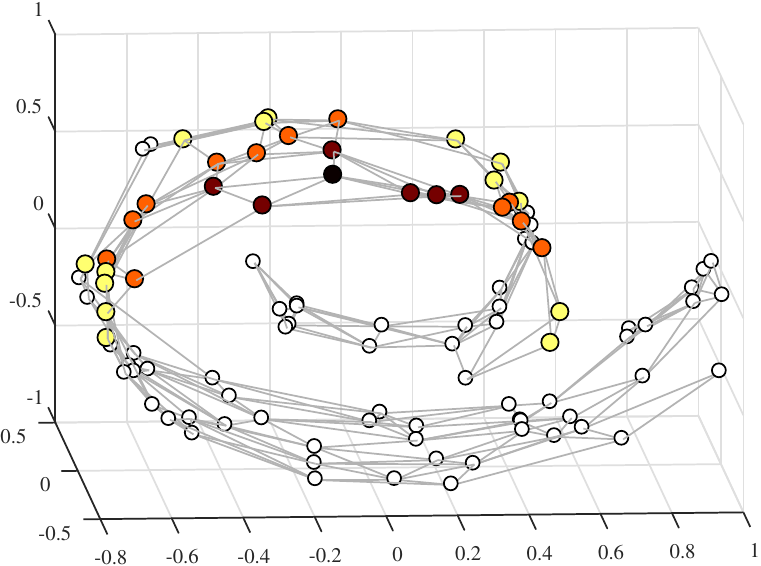} \hspace{6mm} (e)
        \hspace{6mm}
        \includegraphics[scale=0.9]{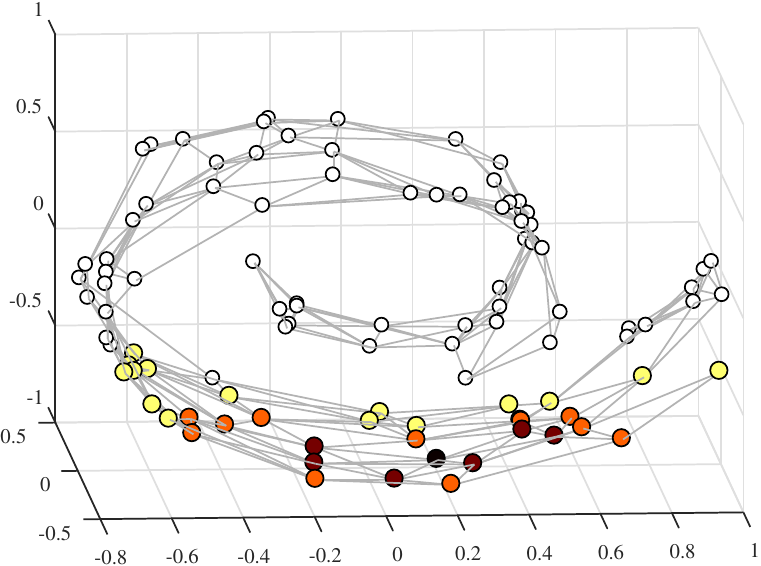} \hspace{8mm} (f)

                $\mathcal{H}_{35,1}(n)=h_{35}(n)u_{1}(n) \sim h_{35}(n)$ \hspace{35mm} 
                $\mathcal{H}_{79,1}(n)=h_{79}(n)u_{1}(n) \sim h_{79}(n)$

        \caption{Localization kernels for vertex-frequency analysis, $\mathcal{H}_{m,k}(n)=h_m(n) u_{k}(n)$, for the case of \textit{vertex domain defined windows} in the local graph Fourier transform, $S(m,k)=\sum_{n=1}^N x(n)\mathcal{H}_{m,k}(n)$. (a) Localization kernel $\mathcal{H}_{28,1}(n)=h_{28}(n)u_{1}(n) \sim h_{28}(n)$, for a constant eigenvector, $u_1(n)=1/\sqrt{N}$, centered at the vertex $m=28$. (b) The same localization kernel as in  (a), but centered at the vertex $m=94$. (c) Localization kernel, $\mathcal{H}_{28,21}(n)=h_{28}(n) u_{21}(n)$, centered at the vertex $m=28$ and frequency shifted by $u_{21}(n)$. Observe kernel amplitude variations, which indicate modulation of the localization window, $h_m(n)$. (d) The same localization kernel as in (c), but centered at the vertex $m=94$. (e) Three-dimensional representation of the kernel $\mathcal{H}_{35,1}(n)=h_{35}(n)u_{1}(n)$, (f) Three-dimensional representation of the kernel $\mathcal{H}_{79,1}(n)=h_{79}(n)u_{1}(n)$.
        }
        \label{VF_windows_neigh}
\end{figure*}

\begin{figure}
        \centering
        \includegraphics[scale=0.9]{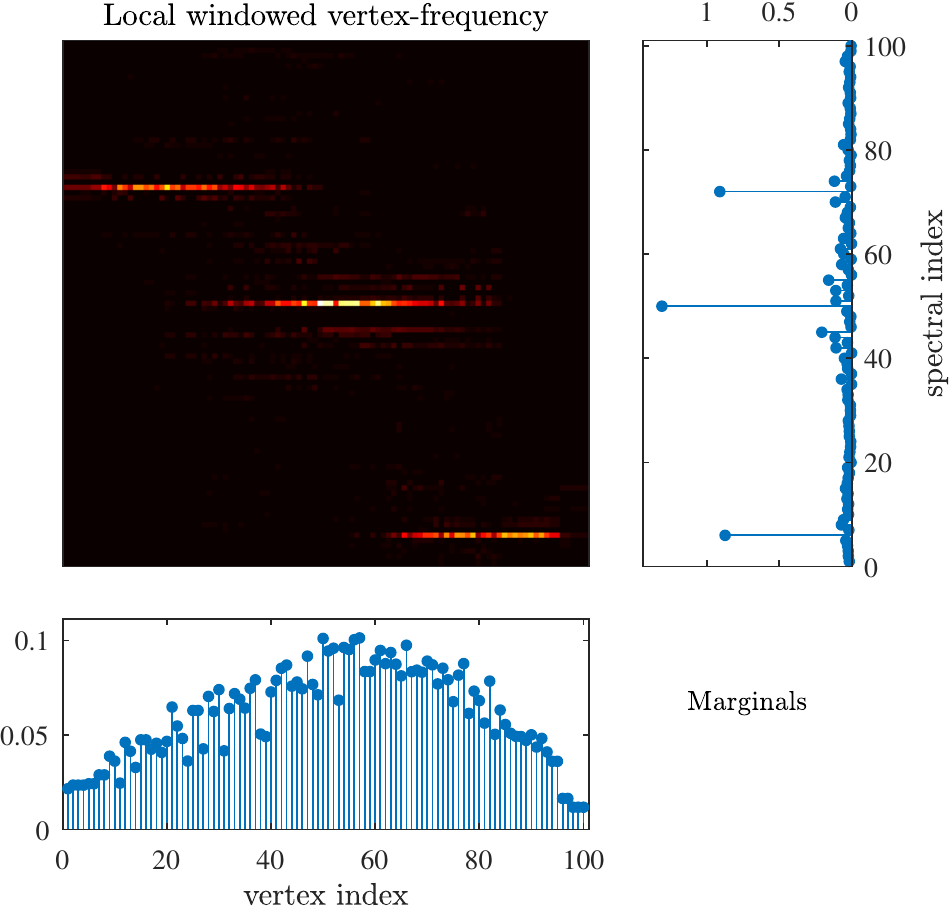}
        \caption{Local vertex-frequency spectrum calculated using the LGFT and the vertex neighborhood windows, as in (\ref{KernelGFDFTmkA}). This representation shows that the graph signal consists of three components located at spectral indices $k=72$, $k=50$, and $k=6$, with  the corresponding vertex indices in their respective vertex subsets $\mathcal{V}_1$, $\mathcal{V}_2$, and $\mathcal{V}_3$, where $\mathcal{V}_1 \cup \mathcal{V}_2 \cup \mathcal{V}_3=\mathcal{V}$. Marginal properties are also given in the panels to the right and below the vertex-frequency representation, and they differ from ideal ones given respectively by $|x(n)|^2$ and $|X(k)|^2$.
        }
        \label{VF_ex1h}
\end{figure}

\end{Example}

\section{Window Parameter Optimization}

The concentration of local vertex spectrum representation can be measured using 
the normalized one-norm \cite{stankovic2001measure}, as 
\begin{equation}
\mathcal{M}=\frac{1}{F}\displaystyle \sum_{m=1}^N\sum_{k=1}^N 
|S(m,k)|=\frac{1}{F} \Vert \mathbf{S} \Vert_1,
\end{equation}
where $F=\Vert \mathbf{S} \Vert_F= \sqrt{ \sum_{m=1}^N\sum_{k=1}^N 
        |S(m,k)|^2}$ is the Frobenius norm of matrix $\mathbf{S}$. Any other norm $\Vert \mathbf{S} \Vert_p^p$ with $0 \le p \le 1$ can be used instead of $\Vert \mathbf{S} \Vert_1$. Recall that norms with $p$ close to $0$ are noise sensitive, while the norm with $p=1$ is the only convex norm, thus allowing a gradient based optimization \cite{stankovic2001measure}.     

\begin{Example}
        The concentration measure $\mathcal{M}(\tau)= \Vert \mathbf{S} \Vert_1 / \Vert \mathbf{S} \Vert_F$ for the signal from Fig. \ref{VF_ex1ab} and the window given in (\ref{win-spect}) and for  various $\tau$ is shown in Fig. \ref{VF_ex1_OPT}, along with the optimal vertex frequency representation. This representation is similar to the one shown in Fig. \ref{VF_ex1g}, where an empirical value of $\tau=3$ was used, with the same localization window and kernel form. 
        
        The optimal $\tau$ can be obtained in  only a few steps through the iteration 
$$\tau_k=\tau_{k-1}-\alpha (\mathcal{M}(\tau_{k-1})-\mathcal{M}(\tau_{k-2})),$$ 
{with $\alpha$ being a step-size parameter.}
\begin{figure}
        \centering
        \hspace*{1cm}\includegraphics[]{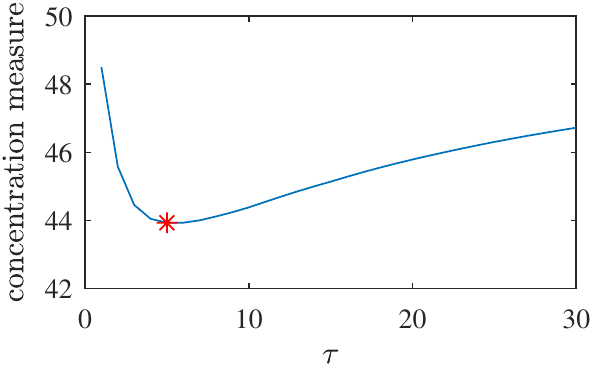}\hspace*{1cm} (a) 
        \vspace{3mm}
        
        \includegraphics[scale=0.9]{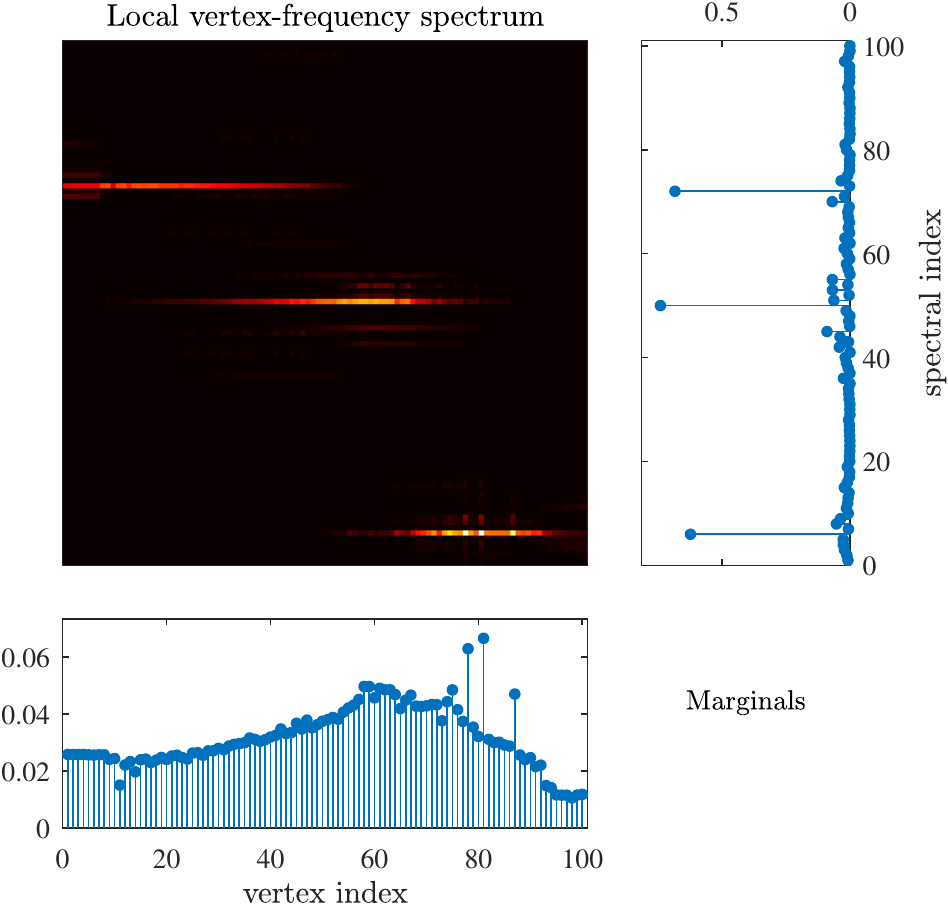}\hspace{-4mm}(b)
        
        \caption{Principle of the optimization of localization window. (a) Measure of the concentration of graph spectrogram for a varying  spectral domain window parameter $\tau$. (b) The corresponding optimal vertex-frequency representation, calculated with $\tau=5$, together  with the marginals.}
        \label{VF_ex1_OPT}
\end{figure}

\end{Example}

The optimization of parameter $\tau$ can also be achieved by using  graph uncertainty principle based techniques \cite{pasdeloup2019uncertainty,Tsitsvero2016,Agaskar}.

\section{Uncertainty Principle of Graph Signals}
In classical signal analysis, the purpose of a window function is  to enhance signal localization in the joint time-frequency domain. However, the uncertainty principle prevents an ideal localization in both time and frequency. Indeed, in the DFT domain the uncertainty principle states that
\begin{equation}
\Vert \mathbf{x}\Vert_0 \Vert \mathbf{X} \Vert_0 \ge N, \label{uncertainityDFT}
\end{equation}
or in other words, that the product of the number of nonzero signal  values, $\Vert \mathbf{x}\Vert_0$, and the number of its nonzero DFT coefficients, $\Vert \mathbf{X}\Vert_0$, is greater or equal than the total number of signal samples $N$; they cannot  simultaneously assume small values. 

To arrive at the \textit{uncertainty principle for graph signals}, consider a graph signal, $\mathbf{x}$, and its spectral transform, $\mathbf{X}$, in a domain of orthonormal basis functions, $u_k(n)$.  Then, the uncertainty principle states that  \cite{pasdeloup2019uncertainty,Tsitsvero2016,Agaskar, elad2001generalized,perraudin2018global}
\begin{equation}
\Vert \mathbf{x}\Vert_0 \Vert \mathbf{X} \Vert_0 \ge \frac{1}{\max_{k,m} \{|u_k(m)|^2\}}. \label{uncertainity}
\end{equation}
This form of the uncertainty principle is generic, and indeed when the basis functions are $u_k(n)=\frac{1}{\sqrt{N}}\exp(j2\pi nk/N)$, the standard DFT uncertainty principle form (\ref{uncertainityDFT}) follows. 

Note, however, that in graph signal processing, the eigenvectors/basis functions can assume quite different forms than in the DFT case. For example, when one vertex is loosely connected with other vertices, then $\max \{|u_k(m)|^2\} \rightarrow 1$  and even $\Vert \mathbf{x}\Vert_0 \Vert \mathbf{X} \Vert_0 \ge 1$ is possible for the condition in (\ref{uncertainity}). This means that a graph signal can be well localized in both the vertex and the spectral domains. 

\begin{Example}
	For the graph shown in Fig. \ref{VF_ex1ab}, we have $\max_{k,m} \{|u_k(m)|^2\} = 0.8565$ which indicates that even $\Vert \mathbf{x}\Vert_0 \Vert \mathbf{X} \Vert_0 \ge 1.1675$ is possible. In other words,  a graph signal for which the number of nonzero samples, $x(n)$, in the vertex domain is just two, will not violate the uncertainty principle even if it has just one nonzero GFT coefficient, $X(k)$. 
\end{Example}

\section{Inversion of the LGFT}\label{GFTInversion}

The inversion relation of the LGFT, calculated using any of the presented localization (window) forms, can be considered in a unified way. Two approaches for the LGFT inversion will be presented next.
\subsection{Inversion by Summation of the LGFT}
 The reconstruction of a signal, $x(n)$, from its local spectrum,  
$S(m,k)$, can be performed through an inverse GFT of (\ref{LGFTDEF}), for the graph windowed signal, $x(n)h_m(n)$,
\begin{equation}
x(n)h_m(n)=\sum_{k=1}^N S(m,k)\, u_{k}(n)
\end{equation}
followed by a summation 
over all vertices, $m$, to yield
\begin{equation}
x(n)=\frac{1}{\sum_{m=1}^{N}h_m(n)} \sum_{m=1}^{N} \sum_{k=1}^{N} S(m,k)u_k(n).
\end{equation}

\begin{Remark}
If the windows, $h_m(n)$, for every vertex, $n$, satisfy the condition
$$\sum_{m=1}^{N}h_m(n)=1,$$ 
then the reconstruction does not depend on the vertex index, $n$, that is the reconstruction is vertex independent.
 In that case 
\begin{equation}
x(n)=\sum_{m=1}^{N} \sum_{k=1}^{N} S(m,k)u_k(n)=\sum_{k=1}^{N} X(k)u_k(n),
\end{equation}
where $X(k)=\sum_{m=1}^{N} S(m,k)$
is a projection of the LGFT onto the spectral index axis. 

For windows obtained using the generalized graph shift in (\ref{AdDEF}), this conditions is always satisfied since $H(1)=1$.
\end{Remark}

The condition  $\sum_{m=1}^{N}h_m(n)=1$ can be enforced by normalizing the elements of the matrix $\mathbf{A}_d$, $d=1,2,\ldots,D-1$ in (\ref{AdDEF}), prior to the calculation of matrix $\mathbf{P}_D$ in such a way that the sum for each column is equal to $1$, to arrive at $$\sum_{m=1}^{N}h_m(n)=\sum_{m=1}^{N}P_D(n,m)=\sum_{d=1}^{D-1}g(d)=const.$$

In general, the local vertex spectrum, $S(m,k)$, can be calculated over a reduced set of vertices, $m \in \mathcal{M} \subset \mathcal{V}$. In this case, the summation over $m$ in the reconstruction formula should be executed over only the vertices $m \in \mathcal{M}$, while vertex-independent reconstruction is achieved if $\sum_{m \in \mathcal{M}}h_m(n)=1$.

\subsection{Inversion of the LGFT with Band-Pass Functions}
For the LGFT, defined in (\ref{PolyVert}) as $
\mathbf{s}_k=\sum_{p=0}^{M-1}h_{p,k} \mathbf{L}^p\mathbf{x}$, the inversion is obtained by a summation over all spectral index shifts, $k$, that is
\begin{equation}
\sum_{k=0}^{K-1}\mathbf{s}_k=\sum_{k=0}^{K-1}\sum_{p=0}^{M-1}h_{p,k} \mathbf{L}^p\mathbf{x}=\sum_{k=0}^{K-1}H_k(\mathbf{L})  \mathbf{x}=\mathbf{x}, \label{unenerINV}
\end{equation}
if
$
\sum_{k=0}^{K-1}H_k(\mathbf{L})=\mathbf{I}. 
$
This condition is equivalent to the following spectral domain condition 
$$\sum_{k=0}^{K-1}H_k(\mathbf{\Lambda})=\mathbf{I}$$ 
since   $\mathbf{U}\sum_{k=0}^{K-1}H_k(\mathbf{\Lambda})\mathbf{U}^T=\mathbf{I}$ and $\mathbf{U}^T\mathbf{U}=\mathbf{I}$. This condition is used to define transfer functions in Fig. \ref{test_bands_a}(a).

\subsection{Kernel-Based LGFT Inversion}
Another approach to the inversion of the local vertex spectrum, $S(m,k)$,  follows the Gabor expansion frameworks \cite{stankovic2014time}, whereby the local vertex spectrum, $S(m,k)$,  is projected back to the vertex-frequency localized kernels, $\mathcal{H}_{m,k}(n)$. The inversion for two forms of the LGFT, defined as in (\ref{LGFTDef1}) and (\ref{FDLGFTLam}), will be analyzed. 

\noindent(a) For the LGFT defined in (\ref{LGFTDef1}), the sum of all of its projections to the localized kernels, $\mathcal{H}_{m,k}(n)$, is
\begin{gather}
\sum_{m=1}^N\sum_{k=1}^N S(m,k)\mathcal{H}_{m,k}(n)=\sum_{m=1}^N \bigg( \sum_{k=1}^NS(m,k)h_m(n)u_k(n) \bigg)
\nonumber \\
=\sum_{m=1}^N \bigg( \sum_{i=1}^N\underset{k \to i}{\operatorname{IGFT}}\{S(m,k)\}\underset{k \to i}{\operatorname{IGFT}}\{h_m(n)u_k(n)\} \bigg) \nonumber \\
=\sum_{m=1}^N\sum_{i=1}^N[x(i)h_m(i)][h_m(n)\delta(n-i)] \nonumber
\\ = \sum_{m=1}^Nx(n)h^2_m(n) 
=x(n)\sum_{m=1}^Nh^2_m(n), \label{GGABINV}
\end{gather}
where IGFT denotes the inverse GFT transform, while  Parseval's theorem for graph signals
$
\sum_{n=1}^{N} x(n)y(n) = \sum_{k=1}^{N} X(k)Y(k) 
$
 was used in the derivation.

The inversion formula for the local vertex spectrum, $S(m,k)$, which yields the original graph signal, $x(n)$, then becomes
\begin{equation}
x(n)=\frac{1}{\sum_{m=1}^Nh^2_m(n)}\sum_{m=1}^N\sum_{k=1}^NS(m,k)h_m(n)u_k(n). \label{GGABINV1}
\end{equation}

\begin{Remark}
This kind of inversion is vertex-invariant if the sum over all vertices $m$ is $n$ invariant and equal to 1, that is
\begin{equation}
\sum_{m=1}^Nh^2_m(n)=1. \label{unener}
\end{equation}
\end{Remark}
        
If the local vertex spectrum, $S(m,k)$, is calculated over a reduced set of vertices, $m \in \mathcal{M} \subset \mathcal{V}$, then the vertex independent reconstruction condition becomes $\sum_{m \in \mathcal{M}}h_m^2(n)=1$.

\smallskip
 
\noindent(b) For the  LGFT with spectral shifted spectral windows, defined by (\ref{FDLGFTLam}) and (\ref{spectral_shifted_kernel}), the kernel based inversion is given by
\begin{gather}
x(n)=\sum_{m=1}^{N}\sum_{k=0}^{K-1}S(m,k)\mathcal{H}_{m,k}(n) \label{PKerLGFTInv}
\end{gather}
if the following condition
\begin{equation}\sum_{k=0}^{K-1}H^2_k(\lambda_p)=1 \label{PKerLGFTInv2} 
\end{equation}
is satisfied for all $\lambda_p$, $p=1,2,\dots,N$.

The inversion formula in (\ref{PKerLGFTInv}), with the condition in  (\ref{PKerLGFTInv2}), follows from
\begin{gather}
\sum_{m=1}^{N}\sum_{k=0}^{K-1}S(m,k)\mathcal{H}_{m,k}(n)\\
=\sum_{m=1}^{N}\sum_{k=0}^{K-1}
\sum_{p=1}^{N} X(p)H_k(\lambda_p)u_p(m) \sum_{l=1}^{N} H_k(\lambda_l)u_l(m)u_l(n) \nonumber.
\end{gather}
Since $\sum_{m=1}^{N}u_p(m)u_l(m)=\delta(p-l)$, the last expression reduces to the graph signal, $x(n)$,
\begin{gather}
\sum_{k=0}^{K-1}
\sum_{p=1}^{N} X(p)H_k(\lambda_p) H_k(\lambda_p)u_p(n)=x(n),
\end{gather}
if the transfer functions, $H_k(\lambda_p)$, $k=0,1,\dots,K-1$, satisfy the condition in (\ref{PKerLGFTInv2}) for all $\lambda_p$.

\subsection{The Wavelet Inversion} 
The wavelet inversion formula
\begin{gather}
x(n)= \sum_{n=1}^N \sum_{i=0}^{K-1} \psi(n,s_i)W(n,s_i)
\end{gather}
can be derived in the same way as in (\ref{PKerLGFTInv}), with the condition in (\ref{PKerLGFTInv2}) assuming the  wavelet transform form
\begin{equation}G^2(\lambda_p)+\sum_{i=1}^{K-1}H^2(s_i\lambda_p)=1, \label{PKerLGFTInv2W} 
\end{equation}
where a set of discrete scales for the wavelet calculation, denoted by $s \in \{s_1, s_2,\dots,s_{K-1}\}$, is assumed with corresponding spectral transfer functions $H(s_i\lambda_p)$, $i=1,2,\dots,K-1$ and a low-pass scale (father wavelet) function $G(\lambda_p)$, playing the role of low-pass function, $H_0(\lambda_p)$, in the LGFT. 

Since the number of wavelet transform coefficients, $W(n,s_i)$, is greater than the number of signal samples $N$, this representation is redundant, and this redundancy allows us to implement the transform trough a fast algorithm, rather than using the explicit computation of all wavelet coefficients \cite{hammond2019spectral,behjat2019spectral,leonardi2013tight}. Indeed, for large graphs, it can be computationally too complex to compute the full eigendecomposition of the graph Laplacian. A common way to avoid this computational burden is to use a polynomial approximation schemes for $H(s_i\lambda)$. One such approach is the truncated Chebyshev polynomial approximations which is based on  the application of the continuous spectral window functions  with  Chebyshev polynomials, which admit order recursive calculation, as in Section \ref{PolyAppo}. If, for a given scale, $s_i$, $i=1,2,\dots,{K-1}$, the wavelet function is approximated by a polynomial in the Laplacian, $P_{i}(\mathbf{L})$ then the wavelet transform can  be efficiently calculated using
\begin{equation}
\mathbf{w}_{i}= P_{i}(\mathbf{L})\mathbf{x},
\end{equation}
where $\mathbf{w}_{i}$ a vector column with elements $W(m,s_i)$, $m=1,2,\dots,N$. Note that this form corresponds to the LGFT form in (\ref{PolyVert}) and (\ref{GenerFunVert}).

\subsection{Vertex-Varying Filtering} 

The filtering in the vertex-frequency domain can be implemented by using the vertex-frequency support function $B(m,k)$. The filtered local  vertex spectrum is then given by
$$
S_f(m,k)=S(m,k) B(m,k)
$$
and the filtered signal, $x_f(n)$, is obtained by the inversion of $S_f(m,k)$ using the above mentioned inversion methods.
The filtering support function, $B(m,k)$, can be obtained, for example, by thresholding noisy values of the local vertex spectrum, $S(m,k)$. 

\begin{Example}
	The inversion relation for the case of band-pass transfer functions and the vertex-varying filtering are verified for the case of signal $x(n)$, from Fig. \ref{VF_ex1ab}(d).
	 %For  better comparison, this signal is shown in Fig. \ref{lgft_band-pass_inversion} (a). 
	 This graph signal is corrupted by an additive white Gaussian noise, at the signal-to-noise ratio of $SNR_{in}=5.5$ dB. 
	 %The corrupted signal is given in \ref{lgft_band-pass_inversion} (b).
	  The LGFT  of the noisy graph signal,  $S(m,k)$, is calculated based on shifted band-pass spectral transfer functions, $H_k(\lambda_p)$, $k=0,1,\dots,K-1$, $p=0,1,\dots,N-1$, of the form defined (\ref{defMeyeWGen}), so that the inversion condition (\ref{PKerLGFTInv2}) holds.  
 The total number of $K=25$ frequency shifted transfer functions, $H_k(\lambda_p)$, of the form as in Fig. \ref{test_bands_a}(b) are used.  A simple thresholding-based filtering support function $B(m,k)=1$ if $|S(m,k)|\ge T$, and $B(m,k)=0$ elsewhere,  is used as the basis for the vertex-varying filtering, $
	S_f(m,k) = \allowbreak S(m,k) B(m,k)
	$, for
	$m=0,1,\dots,N-1$, $k=0,1,\dots, K-1$. The threshold $T=0.078$ was set empirically.	The output graph signal, $x_f(n)$,  is obtained using the inversion relation in (\ref{PKerLGFTInv}) for the filtered LGFT, $S_f(m,k)$. The achieved output SNR was $SNR_{out}=8.94$ dB.  %, and shown in Fig.~\ref{lgft_band-pass_inversion}(c)
	
\end{Example}
A filtering framework for time-varying graph signals may be found in \cite{bohannon2019filtering}. 
%\begin{figure}
%	\centering
%	\includegraphics[]{LGFT_band-pass_win_inversion0SPM}(a)
%	
%	\vspace{2mm} 
%	
%	\includegraphics[]{LGFT_band-pass_win_inversion1SPM}(b)
%	
%	\vspace{2mm}
%	
%	\includegraphics[]{LGFT_band-pass_win_inversion2SPM}(c)
%	
%	\caption{Vertex-varying filtering of a graph signal. (a) The original graph signal, $x(n)$, from Fig. \ref{VF_ex1ab} (d). (b) The graph signal, $x(n)$, corrupted by an additive white Gaussian noise, at $SNR_{in}=5.5$ dB. (c) The resulting graph signal, $x_f(n)$, obtained based on vertex-varying filtering. Filtering is implemented by thresholding the LGFT of noisy graph signal, $S(m,k)$, to produce $SNR_{out}=8.94$ dB.}
%	\label{lgft_band-pass_inversion}
%\end{figure}

\section{Graph Spectrogram and Frames}

Based on (\ref{LGFTDEF}), the \textit{graph spectrogram} is defined as 
\begin{equation}
|S(m,k)|^2=\Big|\sum_{n=1}^N x(n)h_m(n)\; u_{k}(n)\Big|^2.
\end{equation}

Then, according to Parseval's theorem, the \textit{vertex marginal property}, which is a projection of $|S(m,k)|^2$ onto the vertex index axis, is  given by 
\begin{gather*}
\sum_{k=1}^N|S(m,k)|^2=\sum_{k=1}^NS(m,k)\sum_{n=1}^N x(n)h_m(n)\; u_{k}(n) \\
=\sum_{n=1}^N |x(n)h_m(n)|^2,
\end{gather*}
which would be equal to the signal power, $|x(m)|^2$, at the vertex $m$, if $h_m(n)=\delta{(m-n)}$. Since this is not the case, the vertex marginal property of the graph spectrogram is equal to the power of the graph signal in hand, smoothed by the window, $h_m(n)$.

\noindent\textbf{Energy of graph spectrogram.} For the total energy of vertex spectrogram, we consequently have
\begin{equation}
\sum_{m=1}^N\sum_{k=1}^N|S(m,k)|^2=\sum_{n=1}^N \Big(|x(n)|^2\sum_{m=1}^N|h_m(n)|^2\big)\Big).
\end{equation}
If 
$\sum_{m=1}^N|h_m(n)|^2=1$ for all $n$, then the spectrogram on the graph is \textit{energy unbiased} (statistically consistent with respect to the energy), that is
\begin{equation}
\sum_{m=1}^N\sum_{k=1}^N|S(m,k)|^2=\sum_{n=1}^N |x(n)|^2=||\mathbf{x}||^2=E_x.
\end{equation}

\noindent\textbf{The LGFT viewed as a frame.} A set of functions, $S(m,k)$, 
 is called \textit{a frame} for the expansion of a graph signal, $\mathbf{x}$, if \cite{behjat2016signal}
 $$A ||\mathbf{x}||^2 \le \sum_{m=1}^N |S(m,k)|^2 \le B ||\mathbf{x}||^2,$$
where $A$ and $B$ are positive constants. If $A=B$, the frame is termed \textit{Parseval's tight frame} and the signal can be recovered 
as $$x(n)=\frac{1}{A}\sum_{m=1}^N\sum_{k=1}^NS(m,k)h_m(n)u_k(n).$$ The constants $A$ and $B$ govern the numerical stability of recovering the original signal $\mathbf{x}$ from the coefficients $S(m,k)$. 

%The condition for the LGFT, defined as in (\ref{FDLGFTLam}) will be analyzed next.

%\smallskip
%
%\noindent \textbf{(a)} The LGFT, as defined in (\ref{LGFTDef1}), is a frame, since Parseval theorem holds \cite{behjat2016signal,hammond2011wavelets,sakiyama2014oversampled,girault2015stationary}
%\begin{equation}
%\sum_{m=1}^N|h_m(n)|^2=\sum_{k=1}^N|H(k)|^2|u_k(n)|^2,
%\end{equation}
%which allows us to write
%\begin{equation}
%\frac{1}{N}H^2(1)\le \sum_{m=1}^N|h_m(n)|^2 \le \max_{m,k}|u_k(n)|^2\sum_{k=1}^N|H(k)|^2 = \gamma^2 E_h,  \label{FramLGFT}
%\end{equation}
%where $\gamma=\max_{m,k}|u_k(n)|$ and $E_h=\sum_{k=1}^N|H(k)|^2$.
%
%By multiplying each side of inequalities by $||\mathbf{x}||^2$, we get 
%\begin{equation}
%\frac{1}{N}H^2(1)||\mathbf{x}||^2\le \sum_{m=1}^N\sum_{k=1}^N|S(m,k)|^2\le ||\mathbf{x}||^2\gamma^2 E_h.
%\end{equation}
%A frame is termed  \textit{tight frame} if the equality in (\ref{FramLGFT}) holds, that is,  $$\sum_{m=1}^N|h_m(n)|^2=\sum_{k=1}^N|H(k)|^2|u_k(n)|^2=1.$$ 
%The condition is met if $|u_k(n)|^2=1/N$ and $\sum_{k=1}^N|H(k)|^2=1$. 
%
%\smallskip

%\noindent \textbf{(b)} 

The LGFT defined in (\ref{FDLGFTLam}) is Parseval's tight frame if 
\begin{gather}
\sum_{k=0}^{K-1}\sum_{m=1}^N |S(m,k)|^2= 
\sum_{k=0}^{K-1} \sum_{p=1}^{N} |X(p)H_k(\lambda_p)|^2=E_x,\label{TFPKerLGFT}
\end{gather}
where Parseval's theorem for the $S(m,k)$ as the GFT of $X(p)H_k(\lambda_p)$ is used to yield 
$$\sum_{m=1}^N |S(m,k)|^2=\sum_{p=1}^{N} |X(p)H_k(\lambda_p)|^2.$$
 This means that the LGFT in (\ref{FDLGFTLam}) is a tight frame if the condition in (\ref{PKerLGFTInv2}) holds.
This condition is used to define transfer functions in Fig. \ref{test_bands_a}(b) and (c). 

Form (\ref{TFPKerLGFT}), it is easy to conclude that the graph spectrogram energy is bounded with 
\begin{gather}
A E_x\le \sum_{k=0}^{K-1}\sum_{m=1}^N |S(m,k)|^2 \le B E_x,\label{TFPKerLGFTBoun}
\end{gather}
where $A$ and $B$ are respectively the minimum and the maximum of value of $g(\lambda_p)=\sum_{k=0}^{K-1} |H_k(\lambda_p)|^2$. 

\smallskip

 In the same way as in the LGFT case, it can be shown that the wavelet transform also represents a frame with 
\begin{gather}
A||\mathbf{x}||^2 \le \sum_{n=1}^N \sum_{i=0}^{K-1} |W(n,s_i)|^2 \le  B||\mathbf{x}||^2,
\end{gather}
where \cite{hammond2019spectral,behjat2019spectral}
\begin{gather}
A=\min_{0\le\lambda_p\le \lambda_{\max}}{g(\lambda_p)}, \,\,\,\,
B=\max_{0\le\lambda_p\le \lambda_{\max}}{g(\lambda_p)}, \text{\,\,\, and \,\,\,} \nonumber \\
g(\lambda_p)=(G^2(\lambda_p)+\sum_{i=1}^{K-1} H(s_i\lambda_p)|^2), \label{ResFun}
\end{gather}
while the low-pass scale function, $G(\lambda_p)$, is added in the reconstruction formula, $\psi(n,s_0) \rightarrow \psi(n,s_0)+\phi(n,s_0)$, since all $H(s_i\lambda)=0$ for $\lambda=0$. It should be mentioned that the spectral functions  of the wavelet transform, $H(s\lambda_p)$, form Parseval's frame if (\ref{PKerLGFTInv2W}) holds. 

If a continuous (polynomial) approximation of the transfer functions is used, $G^2(\lambda) \approx \bar{P}_{0,M-1}(\lambda)$ and $H(s_i\lambda) \approx \bar{P}_{i,M-1}(\lambda)$, $i=1,2,\dots,K-1$, then an approximation of the constants $A$ and $B$ is obtained finding the respective minimum and maximum of the continuous resulting approximation $g(\lambda)$ in (\ref{ResFun}), within the interval $0\le\lambda_p\le \lambda_{\max}$.

\section{ Vertex-Frequency Energy Distributions}

The energy of a general signal is usually defined as
\begin{gather*}
E=\sum_{n=1}^Nx^2(n)=\sum_{n=1}^Nx(n)\sum_{k=1}^NX(k)u_k(n).
\end{gather*}
This expression can be rearranged as 
\begin{gather*}
E=\sum_{n=1}^N\sum_{k=1}^Nx(n)X(k)u_k(n)=\sum_{n=1}^N\sum_{k=1}^NE(n,k),
\end{gather*}
where for each vertex, the vertex-frequency energy distribution, $E(n,k)$, is defined by \cite{stankovic2018vertex, stankovic2019vertex}
\begin{align}
\!E(n,k)\!\! =\!x(n)X(k)u_k(n) \!\!  =\!\!\! \sum_{m=1}^N \! x(n)x(m)u_k(m)u_k(n) \label{VFE}.
\end{align} 

\begin{Remark}
        The definition in (\ref{VFE}) corresponds to the Rihaczek distribution in classical time-frequency analysis \cite{stankovic2014time,cohen1995time,boashash2015time}. Observe that based on Rihaczek distribution and the expression (\ref{VFE}),  we may obtain a vertex-frequency representation even without a localization window. This very important property  is  also the main advantage (along with the concentration improvement) of classical time-frequency distributions with respect to the spectrogram and STFT based time-frequency representations. 
\end{Remark}

\textit{The marginal properties} of the vertex-frequency energy distribution, $E(n,k)$, are defined as its projections onto the spectral index axis, $k$, and the vertex index axis, $m$, to give 
$$
\sum_{n=1}^NE(n,k)=|X(k)|^2
\quad \text{    and    } \quad
\sum_{k=1}^NE(n,k)=x^2(n).
$$
and correspond respectively to the squared spectra, $|X(k)|^2$, and the signal power, $x^2(n)$, of the graph signal, $x(n)$.   

\begin{Example} Fig. \ref{VF_ex1f} shows the vertex-frequency distribution $E(n,k)$ of the graph signal from Fig. \ref{VF_ex1ab}, together with its marginal properties. The marginal properties are satisfied up to the computer precision, and that the localization of energy is better than in cases obtained with the localization windows in Figs. \ref{VF_ex1g}, \ref{VF_ex1h}, and \ref{VF_ex1_OPT}. Importantly, the distribution, $E(n,k)$,  does not use a localization window.  

\begin{figure}
        \centering
        \includegraphics[scale=0.9]{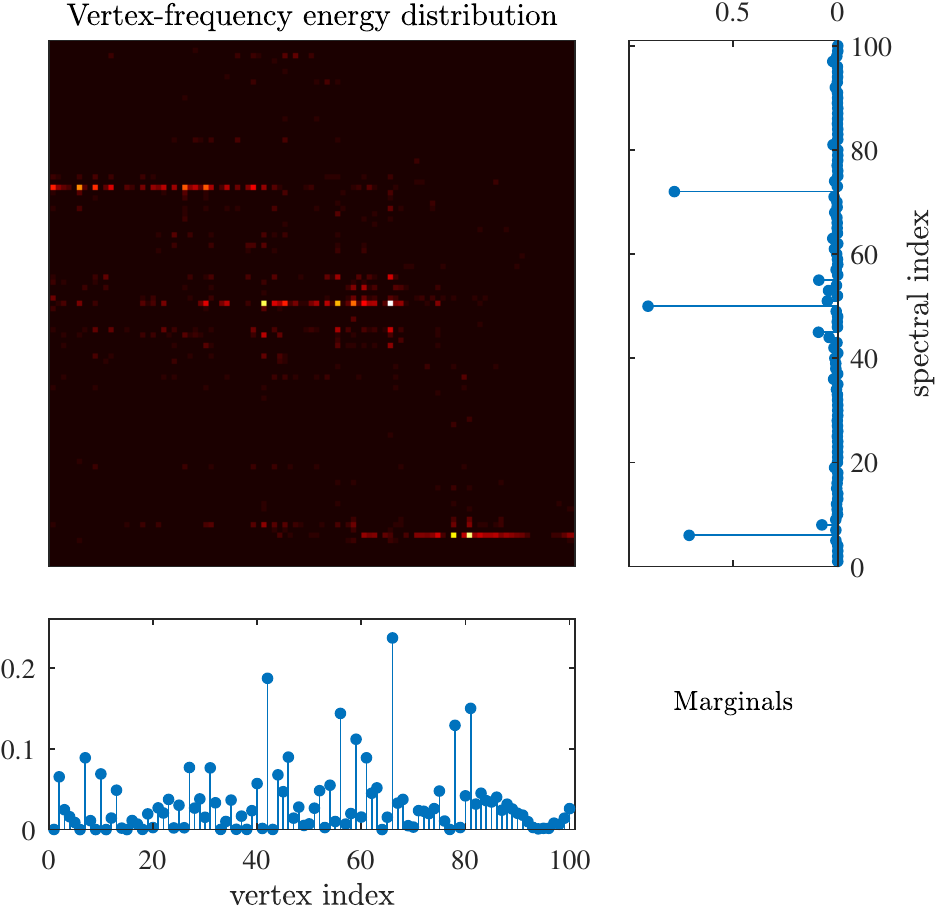}
        \caption{Vertex-frequency energy distribution for a signal whose vertex-frequency representation is given in Fig. \ref {VF_ex1g}. No localization window was used here.}
        \label{VF_ex1f}
\end{figure}
\end{Example}

\subsection{Smoothness Index and Local Smoothness} 

The \textit{ smoothness index} in graph signal processing plays the role of \textit{frequency} in classical spectral analysis, and is defined as the  \textit{Rayleigh quotient} of matrix $\mathbf{L}$ and vector $\mathbf{x}$, that is
\begin{equation}
l=\frac{\mathbf{x}^T\mathbf{L}\mathbf{x}}{\mathbf{x}^T\mathbf{x}} \ge 0. \label{LqfCut}
\end{equation}
\begin{Remark}
	The expression in (\ref{LqfCut}) indicates that the smoothness index can be considered as a measure of the rate of change of a graph signal. \textit{Faster changing signals (corresponding to a high-frequency signals) have larger values of the smoothness index}. The maximally smooth graph signal is a constant signal, $x(n)=c$, whose smoothness index is $l=0$.

In the mathematic literature the inverse of the smoothness index is known as the \textit{curvature} ($\text{curvature} \sim 1/l$). While larger values of the smoothness index correspond to graph signals with larges rates of change (less smooth graph signals), for curvature its larger values would indicate smoother graph signals.   
\end{Remark}

The smoothness index for an eigenvector, $\mathbf{u}_k$, of the graph Laplacian, $\mathbf{L}$,  is equal to its corresponding eigenvalue, $\lambda_k$, that is
\begin{equation}
\frac{\mathbf{u}_k^T\mathbf{L}\mathbf{u}_k}{\mathbf{u}_k^T\mathbf{u}_k}=\lambda_k, \label{LqfCutEV}
\end{equation}
since by definition $\mathbf{L}\mathbf{u}_k= \lambda_k \mathbf{u}_k$.

\begin{Remark}
        If the above eigenvectors are the classical Fourier transform basis functions, then the smoothness index corresponds to the squared frequency of the considered basis function, $\lambda_k \sim \omega^2_k$, while the curvature corresponds to the squared period in harmonic signals.   
\end{Remark}    
  
This makes it possible to define the local smoothness  index, $\lambda(n)$, for a vertex $n$, in analogy with the standard instantaneous frequency, $\omega(t)$, at an instant $t$,   as \cite{dakovic2019local} 
\begin{equation}
\lambda(n)=\frac{\mathcal{L}_x(n)}{x(n)},
\label{local-smooth}
\end{equation}
where it was assumed that $x(n)\ne 0$ and $\mathcal{L}_{x}(n)$ are the elements of the vector $\mathbf{Lx}$.

The properties of the local smoothness include: 

\begin{enumerate}
        
        \item
        The local smoothness index $\lambda(n)$ for a monocomponent signal 
        $$x(n)=\alpha u_{k}(n),$$
        is vertex independent, and is equal to the global smoothness index, $\lambda_k$, since 
        $$\mathcal{L}_{x}(n) = \alpha \mathcal{L}_{u_k}(n)=\alpha \lambda_k u_k(n).$$ 
        In the time-domain signal analysis, this property means that the instantaneous frequency of a sinusoidal signal is equal to its global frequency. 

        \item 
        Assume a piece-wise monocomponent signal 
        $$
        x(n)=\alpha_i u_{k_i}(n)\text{ for } n \in \mathcal{V}_i, \quad i=1,2,\ldots,M,
        $$
        where  $\mathcal{V}_i \subset \mathcal{V}$ are the subsets of the vertices such that $\mathcal{V}_i \cap \mathcal{V}_j=\emptyset$ for $i\ne j$, $\mathcal{V}_1 \cup \mathcal{V}_2 \cup \cdots \cup \mathcal{V}_M=\mathcal{V}$, that is, every vertex belongs to only one subset $\mathcal{V}_i$. Given the monocomponent nature of the signal, within each subset, the considered signal is proportional to the eigenvector $u_{k_i}(n)$.
        
        Then, for each interior vertex $n\in\mathcal{V}_i$, i.e., a vertex whose neighborhood lies in the same set $\mathcal{V}_i$, the local smoothness index is given by
        \begin{equation}
        \lambda(n)=\frac{\alpha_i \mathcal{L}_{u_{k_i}}(n)}{\alpha_i u_{k_i}(n)}=\lambda_{k_i}.
        \label{loc-smooth}
        \end{equation}
        \item
An ideally concentrated vertex-frequency distribution (ideal distribution) can be defined as 
$$I(n,k) \sim |x(n)|^2 \delta\Big(\lambda_k-[\lambda(n)]\Big),$$
whereby it was assumed that the local smoothness index is rounded to the nearest eigenvalue.

This distribution can also be used as a local smoothness estimator, since for each vertex, $n$, the maximum of $I(n,k)$ is positioned at $\lambda_k = \lambda(n)$. The index of the eigenvalue, $\hat{k}$, that corresponds to the local smoothness index  is then obtained as
$$
\hat{k}(n)=\arg \max_k \{I(n,k)\},
$$
so that the estimated local smoothness becomes $\hat{\lambda}(n)=\lambda_{\hat{k}(n)}$.
This estimator is quite common and is widely used in classic time-frequency analysis \cite{stankovic2014time,cohen1995time,boashash2015time}.

\item       

 \noindent\textbf{Local smoothness property.}  The vertex-frequency distribution, $E(n,k)$, satisfies the local smoothness property if  
 \begin{equation}
 \frac{\sum_{k=1}^N \lambda_k E(n,k)}{\sum_{k=1}^N E(n,k)}=\lambda(n).\label{LSPROP}
 \end{equation} 
 In that case, the center of masses of vertex-frequency distribution along the spectral index axis, $k$, should be exactly at $\lambda=\lambda(n)$, and it can be used as an unbiased estimator of this graph signal parameter.  
 
 \end{enumerate}
 
 \begin{Example}
  The vertex-frequency distribution defined by $E(n,k)=x(n)X(k)u_k(n)$ satisfies the local smoothness property. in (\ref{LSPROP}), since 
\begin{gather*}
\frac{  {\sum_{k=1}^N} \lambda_k E(n,k)}{ {\sum_{k=1}^N} E(n,k)}=\frac{ {\sum_{k=1}^N} \lambda_k x(n)X(k)u_k(n)}{ {\sum_{k=1}^N} x(n)X(k)u_k(n)}
=\frac{ \mathcal{L}_{x}(n) }{ x(n)}=\lambda(n).
\end{gather*}
The above relation follows from the fact that $\sum_{k=1}^N \lambda_k X(k)u_k(n)$ are the elements of the IGFT of $\lambda_k X(k)$. Upon employing the matrix form of the IGFT of $\mathbf{\Lambda} \mathbf{X}$, we get  $\mathbf{U}\mathbf{\Lambda} \mathbf{X}= \mathbf{U}\mathbf{\Lambda}  (\mathbf{U}^T \mathbf{U}) \mathbf{X} = (\mathbf{U}\mathbf{\Lambda}  \mathbf{U}^T) (\mathbf{U} \mathbf{X})= \mathbf{L}\mathbf{x}$. With the notation, $\mathcal{L}_{x}(n)$, for the elements of $\mathbf{L}\mathbf{x}$,  we obtain 
$$\sum_{k=1}^N \lambda_k X(k)u_k(n)= \mathcal{L}_{x}(n).$$

        The local smoothness index for the graph signal from Fig. \ref{VF_ex1ab} is presented in Fig. \ref{VF_ex1d}.
        
\begin{figure}
        \centering
        \includegraphics[scale=0.9]{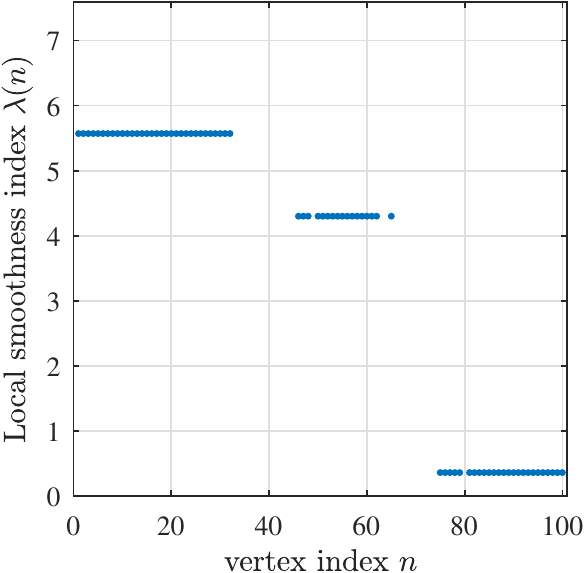}
        \caption{Local smoothness index, $\lambda(n)$, of graph signal from Fig. \ref {VF_ex1ab}. }
        \label{VF_ex1d}
\end{figure}
\end{Example}

\subsection{Reduced Interference Distributions (RID) on Graphs}
{In order to emphasize the relations and the resemblance to the classical time-frequency analysis, in this subsection we will use the complex-sensitive notation for eigenvectors and spectral vectors.} The frequency domain definition of the  energy distribution in (\ref{VFE}) is given by 
\begin{align}
E(n,k) & =x(n)X^*(k)u^*_k(n)  
=\sum_{p=1}^NX(p)X^*(k)u_p(n)u^*_k(n) \nonumber.
\end{align} 
Then, the general form of graph distribution can be defined through introducing a kernel $\phi(p,k,q)$, as \cite{stankovic2018reduced}
\begin{gather}
G(n,k)=\sum_{p=1}^N\sum_{q=1}^NX(p)X^*(q)u_p(n)u^*_q(n)\phi(p,k,q). \label{GENRID}
\end{gather}
Observe that for $\phi(p,k,q)=\delta(q-k)$, the graph Rihaczek distribution in (\ref{VFE}) follows, while the unbiased energy condition $\sum_{k=1}^N\sum_{n=1}^NG(n,k)=E_x$ is satisfied if  $$\sum_{k=1}^N\phi(p,k,p)=1.$$

The so obtained distribution $G(n,k)$ may also satisfy the vertex and frequency marginal properties, as elaborated bellow.
\begin{itemize}
        \item 
        The \textit{vertex marginal property} is satisfied if
        \begin{gather*}
        \sum_{k=1}^N\phi(p,k,q)=1.
        \end{gather*}
        This is obvious from 
        $$\sum_{k=1}^NG(n,k)=\sum_{p=1}^N\sum_{q=1}^NX(p)X^*(q)u_p(n)u^*_q(n)=|x(n)|^2.$$
        
        \item 
        The \textit{frequency marginal property} is satisfied if
        \begin{gather*}
        \phi(p,k,p)=\delta(p-k).
        \end{gather*}
        Then, the sum over vertex index produces  
        \begin{gather*}
        \sum_{n=1}^NG(n,k)=\sum_{p=1}^N|X(p)|^2\phi(p,k,p)=|X(k)|^2,
        \end{gather*} 
        since 
        $\sum_{n=1}^Nu_p(n)u^*_q(n)=\delta(p-q)$, that is, the eigenvectors are orthonormal.
\end{itemize}

\subsection{Reduced Interference Distribution Kernels}

A straightforward extension of classical time-frequency kernels to graph signal processing would naturally be based on exploiting the relation $\lambda \sim \omega^2$, together with an appropriate exponential kernel normalization.  

\begin{figure}
        \centering
        \includegraphics[scale=0.9]{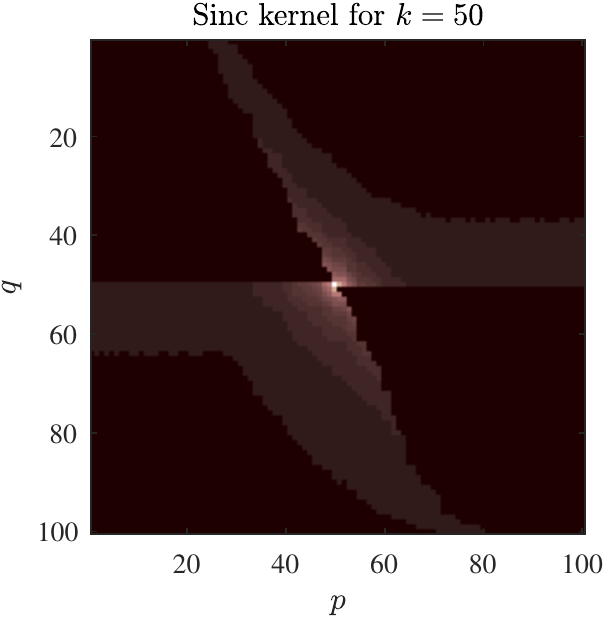}
\caption{The sinc kernel of the reduced interference vertex-frequency distribution in the frequency domain. }
        \label{kerel}
\end{figure}

The simplest reduced interference kernel in the frequency-frequency shift domain,  which would satisfy the marginal properties, is the \textit{sinc kernel}, given by
\begin{gather*}
\phi(p,k,q)=\begin{cases}
\frac{1}{1+2|p-q|}, & \text{for } |k-p| \le |p-q|, \\
0, & \text{otherwise}
\end{cases}
\end{gather*}
and shown in Fig. \ref{kerel} at the frequency shift corresponding to $k=50$. 

\begin{Example}
The sinc kernel was used for a vertex-frequency representation of the signal from Fig. \ref{VF_ex1ab}(d), with the results shown in  Fig. \ref{VF_ex1_RID}. This representation is a smoothed version of the energy vertex-frequency distribution in Fig. \ref{VF_ex1f}, whereby both (vertex and frequency) marginals are preserved. 

\begin{figure}
        \centering
        \includegraphics[scale=0.9]{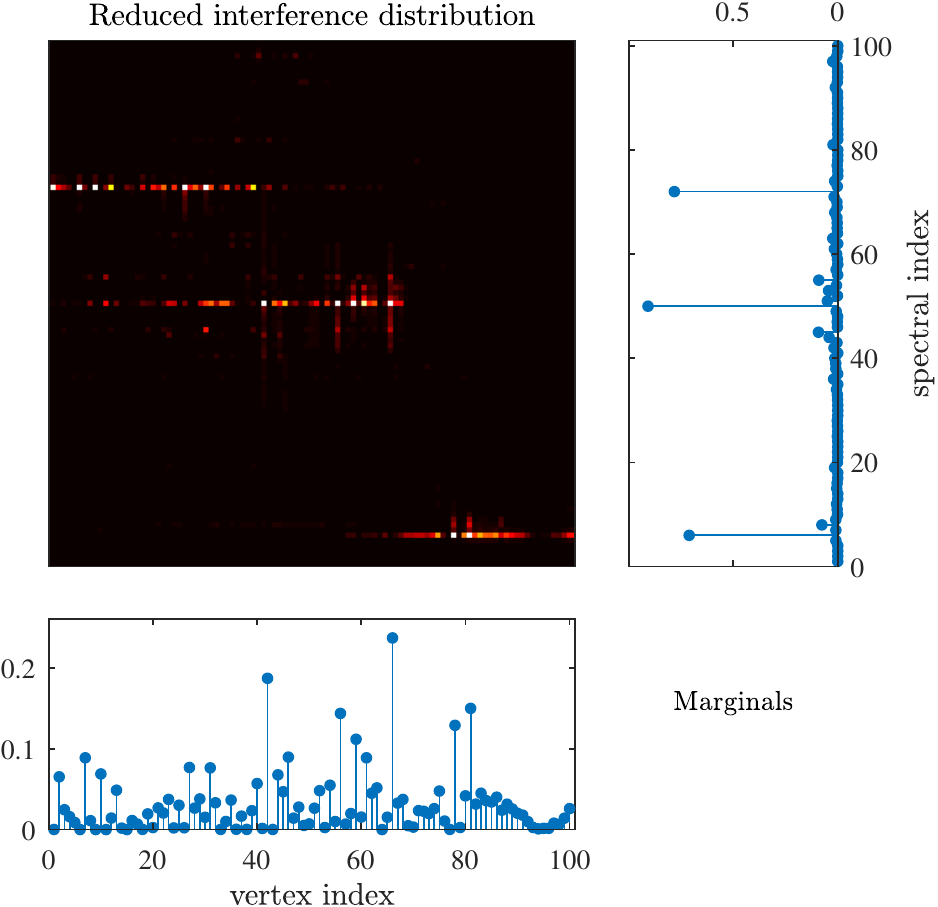}
\caption{Reduced interference vertex-frequency distribution of a signal whose vertex-frequency representation is given in Fig. \ref {VF_ex1g}. The marginal properties are  given in the panels to the right and below the vertex-frequency representation, and they are equal to their corresponding ideal forms given by $|x(n)|^2$ and $|X(k)|^2$. }
        \label{VF_ex1_RID}
\end{figure}
\end{Example}

\begin{Remark} \textbf{ Graph spectrogram and marginal properties.} The general vertex-frequency distribution can be written for the vertex-vertex shift domain as a dual form of (\ref{GENRID})
\begin{gather}
G(n,k)=\sum_{m=1}^N\sum_{l=1}^Nx(m)x^*(l)u_k(m)u^*_k(l)\varphi(m,n,l),
\end{gather}
where $\varphi(m,n,l)$ is the kernel in this domain (the same mathematical form as for the frequency-frequency shift domain kernel).  
The frequency marginal is satisfied if $
\sum_{n=1}^N\varphi(m,n,l)=1$ holds. The vertex marginal is met if $
\varphi(m,n,m)=\delta(m-n).$ The relation of this distribution with the vertex domain spectrogram (\ref{VFSPEC}) is simple, and given by \begin{gather*}\varphi(m,n,l)=h_n(m)h^*_n(l).
\end{gather*}
However, this kernel cannot satisfy both marginal properties, while the unbiased energy condition $\sum_{n=1}^N\varphi(m,n,m)=1$ reduces to (\ref{unener}).
\end{Remark}

\begin{Remark} \textbf{ Classical time-frequency analysis}  follows as a special case from the general form of graph distributions, (\ref{GENRID}),  if the considered graph is a directed circular graph. This becomes obvious upon recalling that the adjacency matrix decomposition produces complex-valued eigenvectors of the form $u_k(n)=\exp(j 2 \pi (n-1)(k-1)/N)/\sqrt{N}$. With 
$$\phi(p,k,q)=\frac{1}{N}\sum_{n=1}^N c(p-q,n)
e^{-j \frac{2\pi (n-1)(k-1)}{N}}e^{j \frac{2\pi (n-1) (p-1)}{N}}$$
	in (\ref{GENRID}),  the classical (Rihaczek based) Cohen class of distributions follows, where $c(k,n)$ is the distribution kernel in the ambiguity domain \cite{stankovic2014time,cohen1995time,boashash2015time}.   
\end{Remark}

\section{Conclusion}
Vertex-frequency analysis, as an approach to the localized analysis of graph signals, is reviewed in this paper. Traditional approaches for graph analysis, clustering and segmentation are based only on the graph topology and spectral properties of graphs. When dealing with signals on graphs, localized analyzes should be focused on data on graphs, incorporating the graph topology. This unified approach to define and implement graph signal localization methods, which takes into account both the data on graph and the corresponding graph topology, is in the core of vertex-frequency analysis. Like in classical time-frequency analysis, the main research efforts are devoted to the graph signal linear representations which include a localization window. Several methods for definition of localization widows in the spectral and vertex domain are presented in this paper. Optimization of the window parameters, uncertainty principle, and inversion methods are also discussed. Following the classical time-frequency analysis, energy forms of vertex-frequency energy and reduced interference distributions, which do not use localization windows, are considered in the second part of the paper. Their role as local smoothness index estimator is presented.

%
%\bibitem{Tsit}
%M. Tsitsvero, S. Barbarossa, and P. Di Lorenzo, ``Signals on graphs: Uncertainty principle and sampling,'' \textit{IEEE Trans. on Signal Processing}, Vol. 64, No. 18, pp. 4845--4860, 2016.
%
%\bibitem{Agas}
%A. Agaskar, and Y. M. Lu, ``A Spectral Graph Uncertainty Principle,'' \textit{IEEE Trans. Information Theory}, Vol. 59, No. 7, pp. 4338--4356, 2013.

%\section{Bibliography}
%\nocite{*}
\bibliographystyle{ieeetr}

\bibliography{graph-signal-processing}

\end{document}